\documentclass[sigconf,authorversion,nonacm]{acmart}
 
\usepackage{amsmath}

\newcommand{\chaptercitation}{%
Roel Vertegaal, Timothy Merritt, Saul Greenberg, Aneesh P. Tarun,
Zhen Li and Zafeiros Fountas. 2026. Interactive Inference:
A Neuromorphic Theory of Human-Computer Interaction. \\
In J. A. Jacko (Ed.), The Human--Computer Interaction Handbook
(4th ed.), CRC Press.%
}

\AtBeginDocument{%
  \providecommand\BibTeX{{%
    \normalfont B\kern-0.5em{\scshape i\kern-0.25em b}\kern-0.8em\TeX}}}

\acmBooktitle{Human Computer Interaction Handbook, Fourth Edition (4th ed.)} 

\usepackage{natbib}
\usepackage{graphicx}
\usepackage{hyperref}
\usepackage{amsmath}
\usepackage{enumitem}
\usepackage{balance}
\usepackage{soul} 
\begin{document}

\title[Interactive Inference: A Neuromorphic Theory of Human-Computer Interaction]{Interactive Inference: A Neuromorphic Theory of Human-Computer Interaction}


\author{Roel Vertegaal}
\affiliation{%
 \institution{Radboud University}
  \country{The Netherlands}
}

\author{Timothy Merritt}
\affiliation{%
  \institution{Aalborg University}
  \country{Denmark}
}

\author{Saul Greenberg}
\affiliation{%
  \institution{University of Calgary}
  \country{Canada}
}

\author{Aneesh P. Tarun}
\affiliation{%
  \institution{Toronto Metropolitan University}
  \country{Canada}
}

\author{Zhen Li}
\affiliation{%
  \institution{Huawei Technologies, Ltd.}
  \country{Canada}
}

\author{Zafeirios Fountas}
\affiliation{%
  \institution{Huawei Technologies, Ltd.}
  \country{UK}
}

\renewcommand{\shortauthors}{Vertegaal et al.}

\begin{abstract}
Neuromorphic Human-Computer Interaction (HCI) is a theoretical approach to designing better user experiences (UX) motivated by advances in the understanding of the neurophysiology of the brain. Inspired by the neuroscientific theory of Active Inference, Interactive Inference is a first example of such an approach. It offers a simplified interpretation of Active Inference that allows designers to more readily apply this theory to design and evaluation. The basic premise in Interactive Inference is that the user predicts a result prior to performing a task. User behaviour is modeled as Bayesian inference on progress and goal distributions that predicts the next action. The difference between the observed result and the prediction is what is processed by the brain. This discrepancy between goal and progress distributions, or Bayesian surprise, can be modeled as a simple mean square error of the signal-to-noise ratio (SNR) of a task. The problem is that the user's capacity to process Bayesian surprise follows the logarithm of this SNR. This means terminal execution errors rise quickly once average capacity is exceeded. Our model allows the quantitative analysis of performance and terminal error using one framework that can provide real-time estimates of the mental load of users that needs to be minimized by design. We show how three basic laws of HCI, Hick’s Law, Fitts’ Law and the Power Law can be expressed using our model. We then test the validity of the model by empirically measuring how well it predicts human performance and terminal error in a car following task. Results suggest that driver processing capacity is indeed a logarithmic function of the SNR of the distance to a lead car, and that terminal error follows the residual surprise predicted by a KL-divergence model. This result provides initial evidence that Interactive Inference can be useful as a new theoretical design tool.

\end{abstract}

\maketitle

\setlength{\footskip}{30pt}

\fancypagestyle{firstpagecitation}{%
  \fancyhf{}
  \renewcommand{\headrulewidth}{0pt}
  \renewcommand{\footrulewidth}{0pt}
  \fancyfoot[C]{%
    \parbox{0.92\textwidth}{%
      \centering
      \scriptsize
      \chaptercitation
    }%
  }
}

\thispagestyle{firstpagecitation}

\section{Introduction}
 There have been many efforts to develop psychological theories of human behaviour as it pertains to user interaction. Frameworks such as GOMS \cite{card:1983} and ACT-R \cite{Anderson:1997ACT-R} have been successful in allowing predictions on time performance when provided with a model user and a model interface. Similarly, models of human error \cite{Reason_1990} have been used to predict error as it relates to the design of interactive systems. However, there has not been a model that has comprehensively explained the relationship between human performance and terminal execution errors in a way that is generalizable. We believe developing such a model is important because it is key to measuring cognitive load of users in real time. Current methods for measuring cognitive load fall short because they either rely on a questionnaire that can be difficult to administer during tasks (e.g., the NASA TLX \cite{Hart:1988TLX}), or on physiological metrics that can be difficult to interpret (e.g., Heart Rate Variability, Electrodermal Activity or Pupillometry). In this chapter, we discuss a theory of human performance as it relates to error that allows \textit{direct measurement} of performance by first defining the signal-to-noise ratio (SNR) of a task. It incorporates the semantics of a task by modeling goals as predictions using Bayesian statistics that are solved using gradient descent. We set out to develop an approach to modeling tasks called Interactive Inference that examines both performance and terminal error using a \textit{relative} information-theoretical perspective. We will show how this approach might allow for wider comparisons of information processing capacity and error of users between otherwise incompatible tasks. While mathematically compatible with Active Inference, we note that our approach simplifies many of its key components and should be viewed as a practical approach rather than a strict implementation of the theory.

\subsection{Key Takeaways}
\begin{enumerate}[leftmargin=*]
\item Interactive Inference Model: We introduce a simple neuromorphic framework called Interactive Inference that integrates human performance and terminal error using information-theoretical constructs. The model uses Bayesian statistics to interpret user goals as predictions, providing a way to measure performance by quantifying the Bayesian surprise as a function of task SNR.

\item Cognitive Load Measurement: The model enables real-time estimation of cognitive load without relying on traditional self-reported questionnaires or physiological metrics. By modeling task complexity externally, it offers a more practical approach for evaluating user experience during interactions.

\item Derivation of HCI Laws: Through known mathematical theorems and logic, we show how foundational principles of HCI — Hick’s Law, Fitts’ Law, and the Power Law of Practice—emerge naturally from our framework through simple methods of gradient descent on a (Bayesian) surprise potential.

\item Empirical Study: To verify whether Interactive Inference could successfully model a task that does not yet have a known model, we demonstrate through an empirical study of a car following task that user performance follows the logarithm of the SNR of the distance to a lead car. 

\item Research Agenda: By discussing the limitations, benefits, and open research questions of this work, we invite further research to extend the theory and explore its practical limitations.

\end{enumerate}

\section{Background}

\subsection{Information Theory and Three Laws of HCI}
Theories of the way the human brain processes information are as old as information theory itself. The very idea that humans might act like a computing circuit arrived shortly after the introduction of the microprocessor. Although most computing logic is deterministic, Shannon \cite{shannon_mathematical_1948} defined self-information as probabilistic: the lower the probability of a message, the higher its information content. One bit of Shannon information is defined as the negative binary logarithm of this probability (negative log probability):
\begin{equation}
\label{equation:SelfInformation}
I(x)=-\log_2(P(x))
\end{equation}
\;
Shannon went on to define that the information capacity $C$ of the transmission of a message through an analog channel has a  maximum that is the (binary) logarithm of the ratio between the signal strength $S$ and the amount of Gaussian noise $N$ in the channel, multiplied by some constant $b$, as expressed by the Shannon-Hartley Theorem:
\begin{equation}
\label{equation:ShannonHartley}
C=b \cdot \log_2 \left( \frac{S+N}{N} \right) = b \cdot \log_2 \left( \frac{S}{N}+1 \right)
\end{equation}
\;

where $b$ is some empirically derived bandwidth parameter. This equation describes the capacity of information sent through the channel in terms of the signal-to-noise ratio. Note that without the negative sign, the SNR represents the reciprocal of the probability of a transmission, the number of distinguishable choices or states. A high signal allows for more uncertainty to be carried, as does a lower noise. This means the very nature of information is statistical, with underlying equations based on (e.g., normal) distributions. This is important because humans, unlike traditional computer algorithms, operate in a probabilistic environment. That is, their exact behaviour cannot be computed, however, it can be modeled using (normal) probability distributions. Using this framework, Card, Moran and Newell \cite{card:1983} developed the Model Human Processor (MHP), a comprehensive set of heuristics for predicting human performance in simple computer tasks. Statistical elements to these models were derived from empirical observations such as those performed by Hick \cite{HicksLaw} and Fitts \cite{Fitts:1954}. They demonstrated that human performance in certain tasks indeed seems to follow a similar pattern to the Shannon-Hartley Theorem: a binary logarithm of the signal-to-noise ratio in the task. E.g., Hick’s Law describes the response time $RT$ for selecting from a number of alternatives $n$, plus 1:
\begin{equation}
\label{equation:Hick}
RT= a + b \cdot \log_2(n+1)
\end{equation}
\;
Here, $a$ and $b$ are some empirically derived constants. $b$ models the time per unit of difficulty (the log component of the task). Response time is derived by multiplying the amount of information in the task by this empirical constant. Fitts \cite{Fitts:1954} derived a very similar law describing the movement time $MT$ of a hand towards a target along a one-dimensional trajectory $A$ with a target width of $W$ (specific equation from \cite{MacKenzie:1992}):
\begin{equation}
\label{equation:Fitts}
MT= a + b \cdot \log_2 \left( \frac{A}{W}+1 \right)
\end{equation}
\noindent One can think of the power of the signal in this equation as the amplitude (one-dimensional distance) to the target, and the noise as the width of the target. Again, $a$ and $b$ are constants that are empirically derived. $b$ describes an average time per bit of difficulty of the task that, when multiplied by the difficulty, gives the movement time prediction. It is important to note the difference between the Shannon-Hartley Theorem and these equations: these equations predict time.

The third law was derived from Snoddy \cite{snoddy:1926} and describes the rate of learning of a model human. Time $T$ to complete a task follows a power law as a function of the number of trials of that task $x$ \cite{Crossman:1959}:
\begin{equation}
\label{equation:PowerLaw1}
T=a \cdot x^{-b}+c
\end{equation}
Again, here, $a$, $b$ and $c$ are empirically determined constants.
\subsection{Predictive Coding and the Bayesian Brain Hypothesis}
One of the drawbacks of classic information-theoretical approaches to human performance modeling is that classic information theory does not describe the presence of memory of messages transmitted through a channel, nor proactive anticipation of those messages. According to Shannon, a bit of information is simply the negative logarithm of the probability of a message (\autoref{equation:SelfInformation}). Only if a message is less probable, does it contain more information. The semantics of how the human interprets that message is not modeled, nor whether the human already knows the message. This is problematic because knowing the message, or anticipating the next message, reduces its information content to zero, as this reduces the uncertainty about the message to zero. 
\begin{figure*}[t]
\includegraphics[scale=0.365]{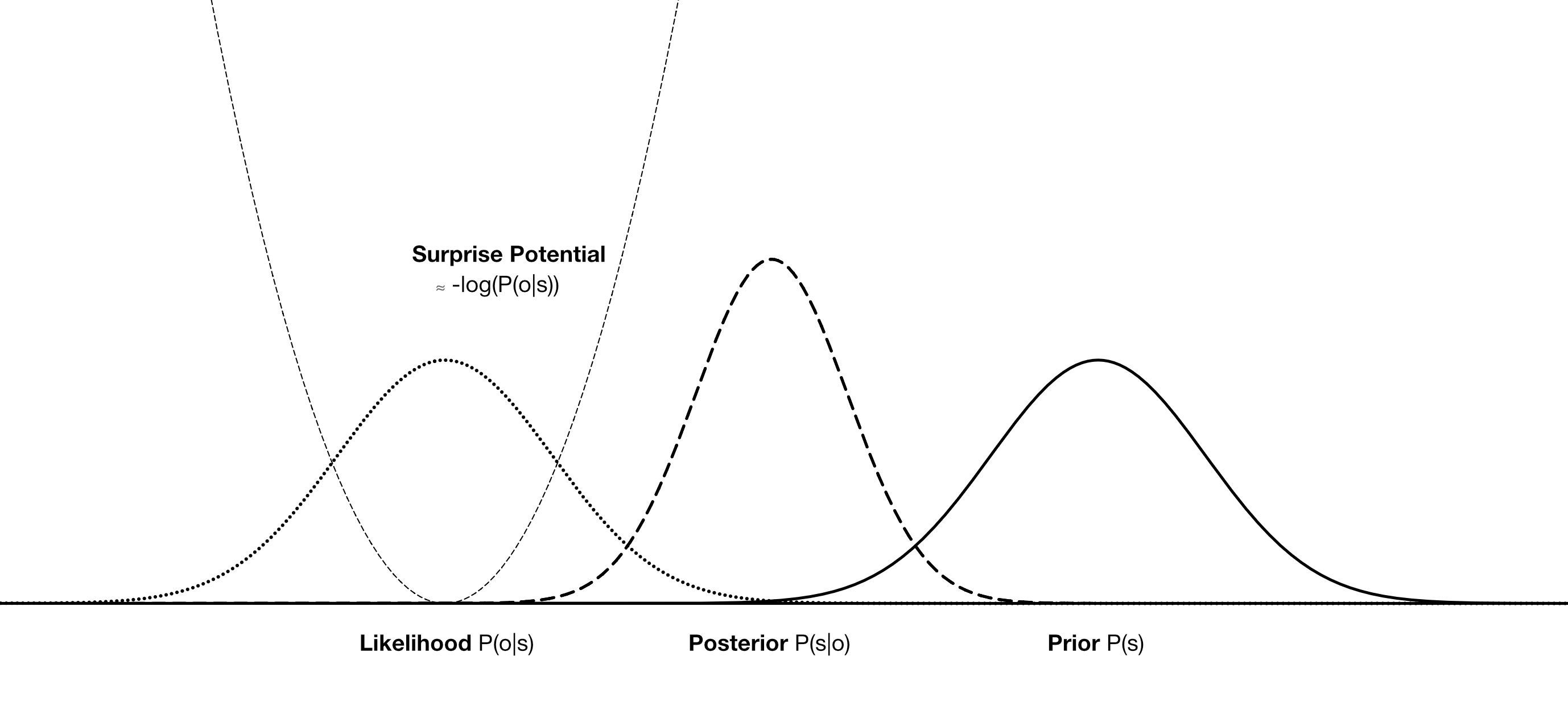}
\caption{\textbf{Bayes Theorem}. Distributions of the likelihood $P(o|s)$ of observations, left, given a prior $P(s)$ that describes the current state, right. In the middle is the calculated posterior $P(s|o)$ that describes the probability of the state given the observation. The quadratic function on the left shows an approximation of the negative log-likelihood, which provides a local surprise potential for updating the prior toward the likelihood and obtaining the posterior during learning. Later, to describe action, the posterior is instead brought toward a goal prior.}
\label{Bayes}
\Description{Distributions of the likelihood P(o|s) of observations, left, given a prior P(s) that describes the current state, right. In the middle is the calculated posterior P(s|o) that describes the probability of the state given the observation. The squared function on the left shows an approximation of the negative log likelihood, the surprise potential that can be used to find the next posterior by gradient descent during learning. Later, to describe action, the posterior is instead brought toward a goal prior.}
\end {figure*}
Instead of framing human processing as a sequential computer processor, McClelland and Rumelhart's parallel processing model \cite{mcclelland:1981} described perception as an interaction of top-down (cognitive/semantic) and bottom-up (sensory) elements. Rao and Ballard \cite{rao:1999} further elucidated this approach by suggesting that the brain builds up statistical knowledge of its environment, i.e., incoming messages, over time, and that this model is used generatively, i.e., top-down, to \textit{predict} the subsequent messages. The error between the prediction and the messages would lead to an update of the prediction. Prediction is convenient for more than one reason: not only does it allow humans to plan actions in their environment based on relevant information, it  serves to compress the information content of their environment by processing only those messages that are surprising: i.e., not, or insufficiently predicted. Predictive coding theory explains how the human brain can work so efficiently and explains the mechanisms behind attention and the scheduling of cognitive resources based on the amount of surprise in the predictions \cite{baldi2010bits}. According to Rao and Ballard \cite{rao:1999}, Bayesian inference \cite{Bayes:63} is a natural candidate for a generative statistical model. Within this context, Bayesian inference estimates the probability that a particular state $s$ (representing a meaning or cause) exists, given an observation $o$. This is called the posterior probability $P(s|o)$, i.e., the chance of state $s$ after, or given observation $o$. It does so by multiplying two probabilities: the likelihood $P(o|s)$ of the observation $o$ given  state $s$, and the $Prior$ probability $P(s)$  of state $s$ \textit{before} the observation. The result is subsequently normalized by dividing by the overall probability of the observation $P(o)$ (see \autoref{Bayes}):

\begin{equation}
\label{equation:BayesTheorem}
P(s|o) = \frac{P(o|s) \cdot P(s)}{P(o)}
\end{equation}
\;

Unlike the frequentist approach to statistics, well-known to the scientific community as hypothesis testing, Bayesian theory thus incorporates prior statistical knowledge in its testing of a hypothesis. The numerator serves as a generative model for predictions. The posterior $P(s|o)$ can be used as the subsequent prior for the next observation and as such, serves to improve the prediction of future observations. This yields a different measure of uncertainty than Shannon information: the difference in information between the prior and posterior, or Bayesian surprise. Because Bayesian surprise incorporates the statistical relationship with prior knowledge, it is semantic, describing the information content of the meaning of an observation under prior knowledge. Bayesian surprise is given by the Kullback-Leibler divergence (KL divergence, \cite{kullback:1997}). It provides the relative entropy between the prior and posterior distributions, or the information gained when updating from the prior distribution $P(s)$ to the posterior distribution $P(s\mid o)$, in bits:

\begin{equation}
\label{equation:KL}
D_{KL} (P(s|o)\:||\:P(s)) = \int P(s|o) \cdot \log_2 \left( \frac{P(s|o)}{P(s)} \right)ds
\end{equation}

\subsection{Entropy and Active Inference}

Friston et al. \cite{Friston2006} further improved upon predictive coding theory by suggesting a mathematical framework for approximating the amount of Bayesian surprise (i.e., relative entropy) processed by the brain. In their framework, called Active Inference, learning and action are unified under a single principle: both use the difference between prior and (an estimate of) the posterior to make decisions. Minimizing the difference informs when to learn or take action. When learning, the difference is minimized by updating the prior in light of the posterior (or if necessary, by altering the likelihood), thus updating the brain's generative model. When taking action, the difference is mainly minimized by altering the world, thereby updating the posterior to be closer to the prior knowledge of what the state of the world should be. 

\subsubsection{Relation to Physical Entropy}
To better understand the concept of relative entropy in information systems, let us use an analogy with the concept of entropy in the physical world. Physical entropy measures the logarithm of the number of microscopic states possible in a system \cite{boltzmann:1866}. Because particle interactions are often random, systems naturally evolve toward states with higher entropy, which can be loosely described as greater disorganization. That is simply because high-entropy states are overwhelmingly more probable. This process is governed by the second law of thermodynamics \cite{Clausius:1879}: in a closed system, entropy (analogous to disorganization) increases over time and approaches equilibrium; it does not spontaneously decrease. Eventually, energy such as heat tends to become uniformly distributed throughout the system. Engines exploit free-energy gradients, which correspond to differences in entropy, to perform work. Random interactions thus explain the natural tendency of heat to dissipate from your morning coffee into the environment. This illustrates the second law of thermodynamics: in closed systems, the entropy, which you can loosely think of as spread, increases. 
Living organisms must counteract this natural tendency towards disorganization by locally staying organized \cite{Friston2012FreeEnergyBiologicalSystems}. That means they maintain their cells in some form of equilibrium called homeostasis. In return, they export entropy to their surroundings through waste products, including heat \cite{Prigogine1977SelfOrganization}. They power this maintenance using \textit{physical} free energy imported from outside the system, from food that concentrates solar energy. Organisms must seek food while avoiding environments that may accelerate entropy increases. For this purpose, they evolved sensorimotor systems and neural circuitry that enable them to observe, learn, and act on their environment. In humans, these systems support homeostasis much like physical systems minimize \textit{physical} free energy: they minimize the variational free energy in their inferences, which bounds prediction error while avoiding needlessly complicated explanations \cite{Friston2012FreeEnergyBiologicalSystems}. This means they learn to make correct predictions about their next state by reducing the Bayesian surprise in their predictions, e.g., of where food might be. Bayesian surprise, then, is the relative entropy between the prior and posterior beliefs about the environment. It encodes the missing information between prediction prior to observation and after observation. The equations governing the statistical properties of particles and Bayesian surprise are thus closely related.

\subsubsection{Action Mirrors Learning}
The term variational refers to a complication in the calculation of Bayesian surprise between prior and posterior: that is estimating the overall probability of observations, the model evidence $P(o)$. For this reason and because the number of known states in the brain is extremely large, Active Inference uses a variational rather than an exact Bayesian approach to calculating the posterior \cite{beal2003variational}. So, Active Inference uses a minimization of free \textit{information} energy (Variational Free Energy, VFE) to reduce uncertainty about its predictions in a way that is mathematically analogous to the Feynman variational principle for physical free energy minimization in statistical mechanics \cite{feynman1972statistical}. It minimizes VFE because it is tractable, and from it an approximation of Bayesian surprise is obtained if an approximated posterior is close to the true posterior. 
In our Interactive Inference interpretation, we try instead to model Bayesian surprise directly, as it is the actual measure of how much our beliefs change when we learn. When learning, sequential Bayesian updates of the prior allow humans to reduce subsequent Bayesian surprise. When acting, similar sequential Bayesian updates of movements occur \cite{Kording:2006}. When learning, the prior is modeled as moving to the posterior, when acting, the posterior as moving to the prior. Once Bayesian surprise is minimal (or rather, stationary), learning or action attenuates and the goal is reached. So, Interactive Inference differs from Active Inference in that we try to avoid approximating VFE, instead focusing directly on Bayesian surprise. 
To briefly summarize Active Inference before we continue, VFE is computed from what is tractable in \autoref{equation:BayesTheorem}: the information in the likelihood function $P(o|s)$ and prior $P(s)$, given by the negative log joint probability of the generative model: 
\begin{equation}
\label{equation:GradientDescent}
-\log_2 P(o,s) = -\log_2 P(o|s) - \log_2 P(s)
\end{equation}
If distributions are normal, this produces a conveniently parabolic error function in which gradient descent \cite{Courant:1943} can be used to approximate the minimum free \textit{information} energy to subsequently approximate Bayesian surprise in a computationally efficient way, while avoiding explicit computation of the information contained in the intractable model evidence $P(o)$. This gradient descent is similar to that used for back propagation in neural networks \cite{Rumelhart:1986}. Unlike classical neural networks, however, VFE allows unsupervised learning because it has an inherent reward or loss function: the minimization of free information energy, which does not require a separate ground truth. In terms of the user’s experience: the reward is inherent when a system performs the function that the user predicted it to perform after some input. VFE is minimized, and Bayesian surprise is correspondingly reduced. In the remainder of this chapter, we will focus on how we might estimate Bayesian surprise more easily and more directly using the externally observable signal-to-noise ratio of a task, while remaining within the general conceptual framework of Active Inference. We emphasize that Interactive Inference simplifies many aspects of Active Inference. Particularly, we do not implement model evidence terms or expected free energy calculations, focusing instead on a tractable framework suitable for HCI applications.

\section{The Concept of “Task” as a Reduction in Relative Entropy}

Interactive Inference implies that the core concept of HCI, the task, can be elegantly modeled as the minimizing of Bayesian surprise between prior knowledge of a desired outcome (the prior or goal) and the posterior (the progress towards a goal given current observations). 

\subsection{Reducing Physical Relative Entropy}
So we have a goal distribution, and a progress distribution, and the brain tries to reduce the relative entropy between them. Let us reconsider how, when we apply this concept to tasks in the real world, this can align with a statistical reduction in the \textit{physical} entropy of the user's environment.

Let us consider a seemingly simple everyday task: organizing socks in a drawer. Because the number of possible places where your socks might end up other than the drawer is infinitely large, random interactions between, e.g., a human and the socks make it much more likely that your socks are anywhere but inside the drawer. The task, then, is to reduce the relative disorganization in the location of the socks, or their physical entropy, relative to the location of the drawer. The goal distribution describes the probability of locations where socks need to be found, inside a drawer. The progress distribution initially describes the probability of the more random locations where socks may be found in the home. We are going to assume both are normal distributions. To execute this task and reduce the physical entropy in the organization of the socks, we must expend physical energy (walk around and pick up socks, place in drawer), i.e., perform work.

According to Interactive Inference, the information used by the user's brain must follow a process proportional to the physical distributions: one that predicts any leftover physical relative entropy between socks and drawer. The brain is merely minimizing the statistical difference in information about the location of socks and drawer that also defines their relative physical entropy, which is in and of itself a statistical property of objects. This means that the brain, in the process of acting to minimize the relative physical entropy is also minimizing the relative information entropy of the task, the Bayesian surprise. To do so, it calculates the statistical likelihood that an observation of a task object corresponds to the outcome of the task, after which it can estimate the chance that the task is complete using Bayesian inference \cite{Friston2006}.

\subsection{Ten Postulates on Capacity and Error in Tasks}
This leads to the first of ten postulates of Interactive Inference:

\begin{enumerate} [itemsep=1ex,leftmargin=*]
\item \emph{Entropy Reduction}. The aim of any \textit{physical} task is to minimize the relative physical entropy in the space of possible task outcomes. Conversely, any errors in the execution of the physical task lead to an increase in relative physical entropy. 
The processing that is required for the user's brain to execute a task leads to a reduction in relative \textit{information} entropy, i.e., a reduction in Bayesian surprise. In our example, this reduction is the \textit{information gain} about the distribution of socks in the household. The aim of the task is to reduce the difference between where the user predicts the socks to be and where they actually are. When executing the task, the user's brain is computing the difference between the probability distribution of locations of socks outside the drawer (likelihood) relative to the probability distribution of the goal (prior): the available locations inside the drawer. In doing so it can, e.g., estimate the time and effort required to complete the task, which leads to postulates 2 and 3:
\;
\item	\emph{Bayesian Surprise}. The relative \textit{information} entropy that needs to be processed by a user to reduce the current error in the task to a satisfactory goal outcome is the Bayesian surprise. Bayesian surprise thus equals the cost, in bits, associated with the accessible work remaining to complete the task given a current error. Bayesian surprise builds up over the scale of task outcomes to form a surprise potential (see \autoref{Bayes}).
\;
\item \emph{Scale of Task Outcomes}. Relative entropy is measured along a scale of possible task outcomes $s$. This scale can be \emph{any} feature of the task space that measures success or failure. In tasks in the physical world relative location (i.e., distance) is often used as a scale of task outcomes. However, an example in the world of information might be a test score. The scale of task outcomes is used to describe the unnormalized error relative to the goal.

\item \emph{Surprise Potential and Difficulty}. The surprise potential is a cost function describing Bayesian surprise at each point along the scale of task outcomes. A fully accomplished task has followed the surprise potential to minimize the unnormalized error on the scale of task outcomes. The surprise potential is often given by the negative logarithm of the probability distribution associated with the goal distribution. If we think in terms of probability distributions, the goal prior indicates that socks should be stored in a drawer. After the task of organizing socks is complete, the probability of obtaining them in the drawer is 100\%, and surprise is thus zero. This means the posterior distribution of socks now corresponds with the goal prior. By reducing relative entropy along the surprise potential, the user thus also reduces the task error, bringing the current location of socks, the \textit{posterior} or \textit{progress distribution $P$}, into correspondence with the memorized \textit{prior}, the \textit{goal distribution $G$}. 

\item \emph{Goal and Progress Distributions}. Goal and progress of a task are defined as probability distributions on the scale of task outcomes $s$. The \textit{goal} distribution $G$ describes the prior, the probability of positive task outcomes on this scale. The \textit{progress} distribution $P$ describes the posterior, the probability of being in a current position along the scale of task outcomes $s$. Each distribution has a mean value $\mu$ on this scale. For the goal distribution, the mean $\mu_g$ signifies the value with the highest probability of a correct task outcome. For the progress distribution, the mean $\mu_p$ denotes the value with the highest probability of the current task outcome. Each distribution also has an associated variance. The standard deviation $\sigma_g$ of the goal distribution defines the base noise scale of the goal. It can be scaled to fit the tolerances of particular instances of the task by a factor $z$. Thus, the actual acceptance region around $\mu_g$ is defined as $z\sigma_g$: completing outside this region constitutes a terminal error. The variance $\sigma_p^2$ of the progress distribution describes the uncertainty with which the task outcome is perceived as moving toward the goal distribution: i.e., the uncertainty in perceived progress on the task. The reciprocal of this variance, $\frac{1}{\sigma_p^2}$, is the precision with which current progress on the task-outcome scale is estimated or controlled. In order for the goal and progress distributions to align, the user must not just reduce the distance between the means of progress and goal distributions to within the z-scaled tolerance $z\sigma_g$, but also bring posterior uncertainty to within the z-scaled tolerance given for the task. A smaller z-scaled tolerance will require smaller updates on the scale of task outcomes as one progresses towards the goal. This is typically achieved not by changing the precision $1/\sigma_g^2$ of the goal distribution, but by following the gradient of the surprise potential. This can result in a halving of the difference between $\mu_p$ and $\mu_g$ in a fixed update time, defining one bit of reduction in the (normalized) task error. A larger z-scaled tolerance allows faster completion time because users do not need to process the least significant bits. The ratio between the difference of means and the base noise scale reflects a normalized error. Although the task tolerance is defined by the z-scaled acceptance region $z\sigma_g$, successful completion requires the progress distribution to become locally compatible with the goal distribution near the accepted region. That is, the posterior uncertainty need not equal the base variance of the goal distribution globally, but near the goal it must be of comparable scale. Under this local approximation, the effective variances of progress and goal may be treated as equal, allowing the negative log probability of the goal distribution \textit{G} to be written as a quadratic surprise potential. This brings us to the next postulate:

\item \emph{Signal and Noise}. We call the distance between the means of the goal and progress distributions on variable $s$ the error \emph{Signal} ($S$) to the user's brain. It describes how far, and in which direction, the user needs to travel along the scale of task outcomes to reach the goal distribution to reduce the error to within the tolerance. The stronger the error signal, the more stimulus the task provides. However, this also means the more accessible work remains in the task. The reciprocal of the error signal represents the accuracy of the task. The standard deviation of the goal distribution defines the task’s base noise scale, specifying the base amount of variation permitted around the goal state. We call this the \emph{Noise} ($N$) parameter of the task, and it relates to the maximum precision required. Thus, the squared reciprocal of this noise scale represents a precision. $N$ can additionally be scaled with a parameter $z$ to define a variable tolerance threshold (confidence interval) across variations of the same task: $zN$.  Returning to the example of organizing socks, the greater the socks’ average distance from the drawer---the error signal $S$---the more difficult the task is to complete. A larger noise scale $N$ does not enlarge the drawer itself; rather, it zooms the entire geometry, so that the same physical discrepancy occupies a smaller normalized distance $S/N$ and therefore represents less accessible work. By contrast, increasing $z$ makes the drawer effectively larger within that scaled geometry, so the socks can be considered put away sooner. The noise scale $N$ and the scaling factor $z$ must therefore be treated as distinct quantities: $N$ sets the geometric scale of the surprise potential, whereas $z$ segments off the acceptance threshold on that scaled potential. From this follows that the relative entropy of a task is a function of the ratio of the mean error $S$ to the noise scale $N$ of the goal as this produces a normalized error scale. Note that this signal-to-noise ratio (SNR) is different from the one introduced by Shannon \cite{shannon_mathematical_1948} in that it describes a normalized error \textit{relative} to a goal. It describes the number of choices, i.e., states, that can be distinguished reliably relative to a goal state:  
\setlength{\abovedisplayskip}{1em}
\setlength{\abovedisplayshortskip}{1em}
\setlength{\belowdisplayskip}{1em}
\setlength{\belowdisplayshortskip}{1em}
\begin{equation}
\frac{\mu_p-\mu_g}{\sigma_g}=\frac{S}{N}
\end{equation}
The ratio $S/N$ defines the normalized task error and represents the accessible work remaining relative to the goal’s noise scale.
\item \emph{KL Divergence}. Relative \textit{information} entropy measures Bayesian surprise, in bits, between progress and goal distributions. It is calculated as the Kullback–Leibler divergence of the progress distribution relative to the goal distribution. The logarithm expresses the subtraction of the two log probabilities in bits, while the integral averages this quantity over the progress distribution:
\begin{equation}
\label{equation:KL2}
D_{KL} (P(s)\:||\:G(s)) = \int P(s) \cdot \log_2 \left( \frac{P(s)}{G(s)} \right)ds
\end{equation}
\item \emph{Equal Variance Condition}. The KL divergence can be greatly simplified in cases where the goal and progress distributions are normal, with identical variance. Bayesian surprise in bits can be calculated using a simplified form of the KL divergence: approximately half the squared signal-to-noise ratio, where the signal is the distance between the means of the progress and goal distributions and the noise is the standard deviation of the goal distribution. The mathematical proof for the relative entropy of two Gaussians with equal variance is the motivation for using squared error in statistics and machine learning \cite{Novak:2007}:
\begin{equation}
\label{equation:KL_SNRderivation}
  \begin{split}
\frac{P(s)}{G(s)} = \frac{\sqrt{\sigma_{g}^{2}}}{\sqrt{\sigma_{p}^{2}}}\exp\left(-\frac{(s-\mu_{p})^{2}}{2\sigma_{p}^{2}}+\frac{(s-\mu_{g})^{2}}{2\sigma_{g}^{2}}\right) \text{ and } 
 \sigma_p =\sigma_g
\\
\log_{2}\frac{P(s)}{G(s)}=\log_{2}\left(\frac{\sigma_{g}} {\sigma_{g}}\right)+\log_{2}\exp\left(-\frac{(s-\mu_{p})^{2}}{2\sigma_{g}^{2}}+\frac{(s-\mu_{g})^{2}}{2\sigma_{g}^{2}}\right) 
\end{split}
\end{equation}
The integral of Gaussian $P(s)$ multiplied by a function $log_2 \left( \frac{P(s)}{G(s)} \right)$ equals the expectation under $P(s)$:
\begin{equation}
D_{KL} \left(P(s)\:||\:G(s)\right)
= \mathbb{E}_p \left[ \frac{-\left(s-\mu_p\right)^{2}+\left(s-\mu_g\right)^{2}}{2\ln\left(2\right)\sigma_g^{2}}\right] 
\end{equation}
The expectation under $P(s)$ equals:
\begin{equation}
\begin{split}
\mathbb{E}_p\left[(s-a)^{2}\right]=(\mu_p-a)^{2}+\sigma_p^{2}
\\
\mathbb{E}_p \left[ -\left(s-\mu_p\right)^{2} \right]=-\left((\mu_p-\mu_p)^{2}+\sigma_p^{2} \right)= -\sigma_p^2
\\
\mathbb{E}_p \left[ \left(s-\mu_g\right)^{2} \right]=(\mu_p-\mu_g)^{2}+\sigma_p^{2}
\end{split}
\end{equation}
Substituting expectations:
\begin{equation}
=\frac{\left( \mu_p-\mu_g \right)^2 +\sigma_p^{2}-\sigma_p^{2}}{2\ln\left(2\right)\sigma_g^{2}}
\end{equation}
\;

For Gaussian distributions with equal variance, the KL divergence in bits is:
\begin{equation}
\label{equation:KL_SNR}
D_{\mathrm{KL}}\!\left(P(s)\,\|\,G(s)\right)
=
\frac{1}{2\ln2}
\left(
\frac{\mu_p-\mu_g}{\sigma_g}
\right)^2
=
\frac{1}{2\ln2}
\left(
\frac{S}{N}
\right)^2
\end{equation}

We note that the metric $\frac{\mu_p-\mu_g}{\sigma_g}$ is mathematically identical to Cohen's d \cite{cohen1988statistical}, a well-established statistical measure of effect size between two distributions. The resulting quadratic form defines the canonical surprise potential. It is equivalent to the potential energy of a Hookean spring, where the error signal is the displacement from equilibrium and $1/N^2$ the stiffness constant \cite{feynman1972statistical}. Conveniently, it is also equivalent to the negative log of a Gaussian after dropping the normalization constant. The quadratic function can therefore be interpreted as a surprise potential generated by an unnormalized Gaussian distribution.
\;

\item \emph{Performance}. Human performance, or capacity to process relative entropy, follows the \textit{reduction} in KL divergence as the user moves along the scale of task outcomes until the threshold tolerance \textit{zN} is reached. This capacity to process Bayesian surprise, in bits, follows Shannon-Hartley's theorem (\autoref{equation:ShannonHartley}) but with reduction in relative, rather than absolute entropy. This information gain equals the logarithm of a z-scaled signal-to-noise ratio plus 1: 
\begin{equation}
\label{equation:Capacity}
C = b \cdot \log_2 \left( \frac{\mu_p-\mu_g}{z\sigma_g} +1 \right) = b \cdot \log_2 \left( \frac{S}{zN} + 1 \right)
\end{equation}
\;
where \(C\) is the information-processing capacity, measured in bits; \(log_2\!\left(\frac{S}{zN}+1\right)\) is task difficulty in bits; and \(b\) is the empirically determined mean capacity coefficient. Task difficulty is the logarithmic imprint of the surprise potential that guides the reduction of normalized error. Human performance and error can thus both be described on the same scale of signal-to-noise ratio (SNR). Note that this is a different scale than the task outcome scale. It describes the ratio of the error between means and the standard deviation of the goal, i.e., the normalized error on the task outcome scale. Since performance is a logarithmic function of the SNR, while the amount of Bayesian surprise in a task is a squared function of SNR, users will run out of capacity to process Bayesian surprise as the SNR increases. The user's capacity to process Bayesian surprise can be measured, in real time, by calculating the average reduction in KL divergence per unit change in normalized task error. This capacity corresponds to the cognitive and/or sensorimotor load induced by a task because it quantifies the relative amount of Bayesian surprise processed by the brain. As per Shannon-Hartley's theorem \cite{shannon_mathematical_1948}, maximum error-free performance is defined by the slope or mean capacity coefficient of the logarithmic function on the normalized error scale of SNR. We can express the normalized scale directly in bits if we apply a binary logarithm, in which case capacity becomes a linear function of task difficulty. 

\item	\emph{Error}. Terminal error—finishing a task outside the tolerance threshold—is modeled as residual Bayesian surprise after the spare capacity, defined as the maximum error-free performance minus current capacity ($b-b\log_2\!\left(\frac{S}{zN}+1\right)$), reaches zero. Once capacity is exhausted, terminal error can no longer be kept arbitrarily low \cite{shannon_mathematical_1948}; instead, it follows an empirical KL-divergence form that increases as a square function of SNR:
\begin{equation}
\frac{\beta}{2\ln 2}\left(\frac{S}{N}\right)^2
\end{equation}
where $\beta$ is an empirically determined coefficient relating the residual surprise to a squared normalized error. This error surprise can be measured and predicted in bits, using the same normalized error scale of SNR. It can be empirically modeled by calculating the surprise of \textit{not} having an error, the negative log probability of the complement of the normalized frequency of errors per unit of SNR: 
\begin{equation}
\label{equation:KL_SNR2}
E_b\equiv-\log_2\left(1-p_{\mathrm{err}}\left(\frac{S}{N}\right)\right), \; \widehat{E}_b = \frac{\hat{\beta}}{2\ln 2}\left(\frac{S}{N}\right)^2 
\end{equation}

Here, $p_{\mathrm{err}}(S/N)$ denotes the empirical error probability at each level or bin of normalized error, regressed against the KL divergence.
\end{enumerate}
\noindent The ultimate purpose of any user interface design, then, is to allow the user to achieve Postulate 1 while avoiding Postulate 10. Together, our postulates give us a mathematical tool grounded in both physics and neuroscience that allows us to measure the capacity of the user as it relates to potential error, in real time. 
\begin{figure}[t]
\includegraphics[width=\columnwidth,trim={1cm 2cm 1cm 1.1cm},clip]{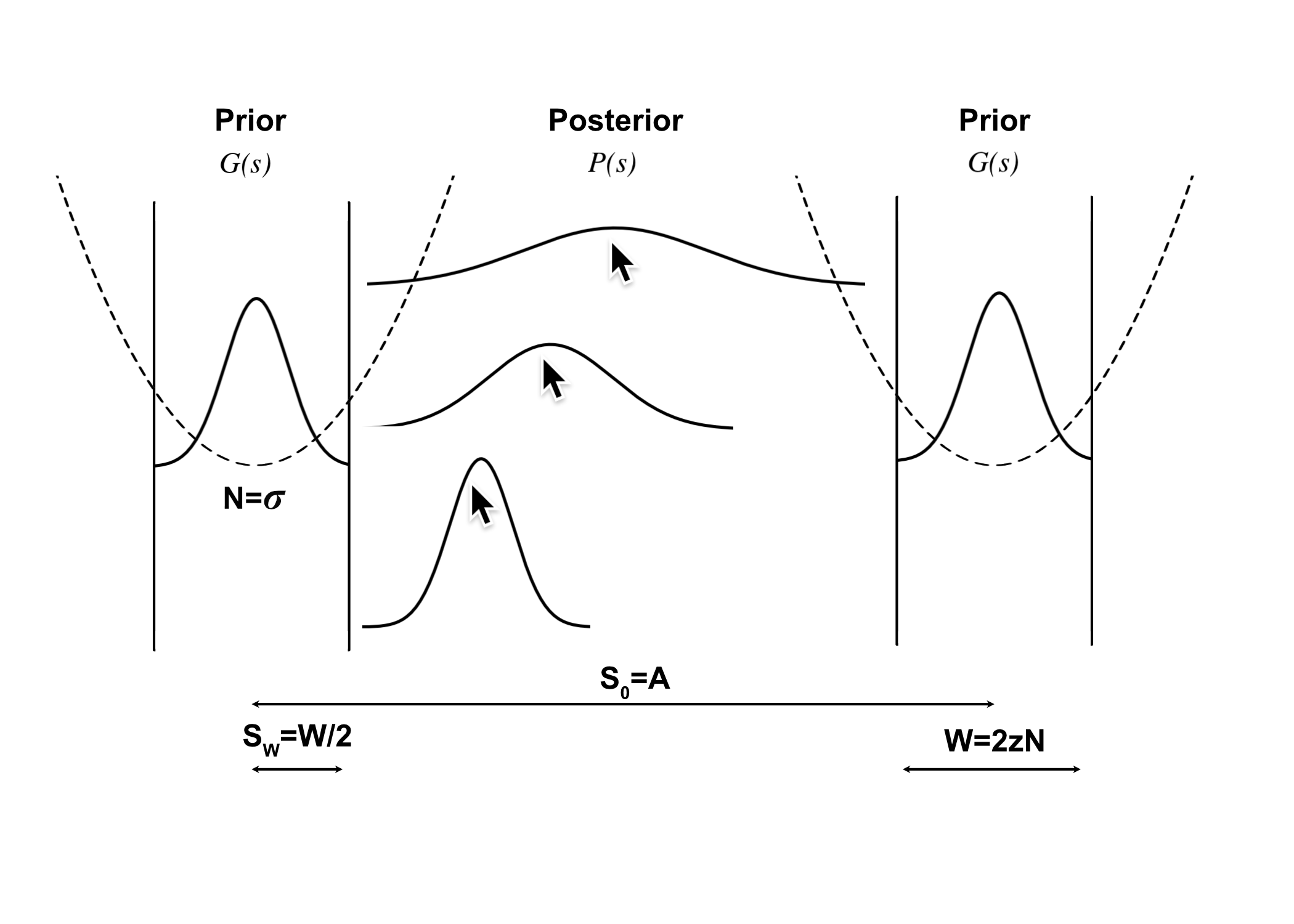}
\caption{\textbf{Bayesian Fitts' Law}. Progress distribution $P(s)$ (the posterior probability of the cursor being within tolerance) from top to bottom: in the middle of a trial, at 2/3 s of the trial and at the end of a trial. Tolerance is the $S_w=zN$ interval on the goal distribution $G(s)$ (the prior), on both sides for a reciprocal trial. Cursor acceleration follows from the gradient of the surprise potentials of the goals (dashed), which is defined as the negative log probability of observing an error S under noise scale N.}
\label{Fig:Fitts}
\Description{Progress distribution $P(s)$, the posterior probability of the cursor being within tolerance, from top to bottom: in the middle of a trial, at 2/3 s of the trial and at the end of a trial. Tolerance is the $S_w=zN$ interval on the goal distribution $G(s)$ (the prior), on both sides for a reciprocal trial. Cursor acceleration follows from the gradient of the surprise potentials of the goals (dashed), which is defined as the negative log probability of observing an error S under noise scale N.}
\end {figure}
\section{Three Laws of HCI Revisited}
The three laws of Human-Computer Interaction can be recast in terms of the above equations. These include Fitts’ Law, Hick's Law, the Power Law of Practice and, potentially, others that are yet to be discovered using our method.

\subsection{Fitts' Law}
Kording and Wolpert's work \cite{Kording:2006,Wolpert:2007} strongly suggests that sensorimotor control is indeed guided by Bayesian prediction. Wu et al. \cite{wu:2006} and Williamson et al. \cite{Williamson_2022} also suggest that Fitts' Law \cite{Fitts:1954,MacKenzie:1992} is the result of the sensorimotor system performing Bayesian inference on movement. \autoref{Fig:Fitts} illustrates this process. A Fitts' Law task involves moving the cursor (or hand) reciprocally between two goals at distance $A$ and click within width $W$. Fitts' Law experiments systematically vary both the distance $A$ between goal means and the tolerance $W$ around each mean \cite{MacKenzie:1992}. The goal can be modeled as a probability distribution $G(s)$ with base noise scale $N$. The target width \(W\) defines an acceptance interval $S_W=zN$ on both sides of $G(s)$'s mean, such that $W=2zN$. The figure also shows a cursor. Cursor velocity reflects the uncertainty of the posterior progress estimate $P(s)$ relative to the goal $G(s)$. As the cursor moves, velocity, and thus uncertainty in the posterior, increases to a maximum, after which it decreases to a minimal velocity or uncertainty once the target is entered. Prior to that, velocity and thus uncertainty in the movement must fall within tolerance $zN$ so as not to click outside the target. In Interactive Inference, then, movement is modeled by a step-wise iterative update of the posterior toward the prior(s). 

The initial error signal $S_0$ is equal to $A$. During each Bayesian update the brain then estimates new velocities based on the derivative or slope in the segment of the surprise potential delineated by $S_0$ and $S_W$ ($\frac{W}{2}$). This surprise potential is approximately the negative log probability of the goal distribution, and equals the KL divergence in \autoref{equation:KL_SNR}. The noise scale $N$ is a fixed precision estimate that does not change with $W$. Indeed, Welford suggested early on that performance is a function of a signal-to-noise ratio in the brain and that there is a maximum capacity at which the brain can process a Fitts' Law task due to the \textit{``neural noise blurring or distorting signals''}~\cite{Welford_fundamentals_1968}. This describes the base noise scale $N$. 

MacKenzie \cite{MacKenzie:1992} suggested the following equation for describing the resulting movement time in a Fitts' Law task:
\begin{equation}
\label{equation:FittsLaw2}
MT= a + b \cdot \log_2 \left( \frac{A}{W} + 1 \right)
\end{equation}
\;
\noindent Here, $MT$ equals movement time, and $a$ and $b$ are empirically determined constants. The equation is modeled after the Shannon-Hartley theorem in \autoref{equation:ShannonHartley}. We can instead define a surprise potential $U$ in the form of \autoref{equation:KL_SNR} that can be minimized over time:
\begin{equation}
\label{equation:FittsLaw3}
s = \frac{S}{N}, \ \   U=\frac{1}{2} s^2
\end{equation}
\;

where the initial condition $S_0 = A$, and $1/N^2$ represents the maximum spatial motor precision $1/\sigma_g^2$ required for the cursor or hand movement in the task, thus defining an invariant geometric scale. This surprise potential has the same form as the potential energy of a spring. If we therefore express the movement as a first-order relaxation of this “spring” (i.e., the derivative of \autoref{equation:FittsLaw3}), we obtain:
\begin{equation}
\label{equation:affine}
\frac{\partial U}{\partial s} = s, \ \ 
\dot{s}=-\frac{1}{b'}s
\end{equation}
\;

\noindent Here, $\frac{1}{b'}$ governs the \textit{temporal relaxation rate} or temporal precision at which the system relaxes along the surprise potential over time. It is proportional to the index of performance, the reciprocal of $b$ in \autoref{equation:FittsLaw2}. The stop condition is when the cursor finds the near edge $S_W \leq zN$. The resulting behaviour corresponds to a binary search for the target in which the error in the cursor position is controlled by halving the distance to the target, at each time interval, until the near edge is found. Each halving can be considered a bit. \autoref{equation:FittsLaw3} describes the control function that encodes the potential energy underlying the planning of the cursor (or hand) movement. \autoref{equation:affine} models the corresponding force field and its first order relaxation over time. Modeling actual movement, however, requires converting the surprise potential into kinetic motion. This can be achieved by using a second-order gradient-descent model in which the gradient of the potential determines cursor acceleration rather than velocity. This accounts for the inertia of the input device and limb, represented by mass, and for viscoelastic resistance of muscles and input device, represented by damping. A full treatment of second-order dynamics is beyond the scope of this chapter. Regardless, in a reciprocal tapping task between the centers of targets, the relaxation of potential energy over time produces the logarithmic time result found in \autoref{equation:FittsLaw2} \cite{Knuth1998TAOCP3}:
\begin{equation}
\label{equation:FittsLaw4}
S(t)=S_0 e^{-t/b'}
\;\;\Longrightarrow\;\;
MT=b'\ln\!\left(\frac{S_0}{S_W}\right)
=
b'\ln 2\log_2\!\left(\frac{S_0}{S_W}\right)\end{equation}
\;
Here, $S(t)$ is the time-based error signal in this task, with $S_0$ the initial distance between the center of the goals. $S_W$ describes the stop condition in this task ($\frac{W}{2}$). \autoref{equation:FittsLaw4} yields the basic Fitts' Law form \cite{Fitts:1954}. A one may be added to the logarithmic argument to account for the null choice (where the hand is already inside the target) \cite{MacKenzie:1992}.
Empirical validation of the variational Bayesian underpinning of Fitts' Law remains ongoing. However, Schmidt et al. \cite{schmidt1979motor} showed the variance in motor impulse in a similar task to be linearly related to the variance in the resulting distribution. 

Our surprise formulation can also be used to predict terminal error, if the normalized frequency of such errors is converted to bits using negative log probability. From our earlier discussion, it follows that once a user runs out of capacity, the surprise associated with not making a terminal error---defined as the negative logarithm of the complement of the normalized frequency of terminal errors---should have the same quadratic dependence on normalized error as the equal-variance KL divergence in \autoref{equation:KL_SNR2}. There is some evidence for such \textit{error surprise} curve \cite{LoechesDeLaFuente2014}. Welford \cite{Welford_fundamentals_1968} demonstrated a method for post hoc correction of goal distributions, so as to adjust for differences in terminal error rates between different devices. His work appears consistent with the above discussion. Wobbrock et al. \cite{wobbrock2008error} also modeled the probability of a terminal error, with good fit. Our model predicts that error surprise only tracks the KL divergence once capacity is reached, while their model describes the probability of terminal error rather than the negative log probability. Guiard et al. \cite{guiard2011fitt} analyzed the terminal error in different conditions of a Fitts' Law experiment. Their RVE metric also appears to be consistent with \autoref{equation:KL_SNR2}.

\subsection{Hick-Hyman Law}
Hick's Law, better known as Hick-Hyman Law \cite{HicksLaw,Hyman1953StimulusIA} can also be derived from our surprise potential. Hick-Hyman Law describes user response time when the task is to select one from a number of choices. Hicks \cite{HicksLaw} found that the response time was a logarithm of the number of choices $n+1$:
\begin{equation}
\label{equation:Hicks}
RT=a+b \cdot \log_2(n+1)
\end{equation}
Here, $a$ and $b$ are some empirically derived time constants. $b$ can be used to convert the difficulty of the task to a time estimate. Unlike Fitts' Law, we model Hick-Hyman Law as a learning task in which the user performs Bayesian inference to update a prior of possible targets with an observation of a single highlighted candidate to create a posterior of only one candidate. This means the role of the prior and posterior distributions are not reversed: in this task we update the prior to the posterior. Hyman \cite{Hyman1953StimulusIA} extended Hick's experiment to include unequal probabilities for each choice. He found that time results were a function of the entropy of the task:
\begin{equation}
\label{equation:HicksHyman}
\sum_{i=1}^{n}p_{i}\log_{2}\left(\frac{1}{p_{i}}\right)
\end{equation}
\;
where $p_i$ is the probability of each choice in the set. Note that as the set gets smaller, the chance of a correct answer increases. As the set gets larger, it decreases by the number of elements in the set. This means the only relevant parameter in this law is $p_i$. According to Wu et al. \cite{wu:2018Hick}, evidence from MRI scans of the brain indeed suggests that the narrowing of the selection from the prior to the observation follows step-wise Bayesian posterior updates. According to them, the cognitive control network in the  brain eliminates half the options at each Bayesian update, yielding a binary search with a logarithmic response time. Within our framework, Hick-Hyman law can be rewritten using the simple first-order gradient descent in \autoref{equation:affine} on the canonical surprise potential from \autoref{equation:FittsLaw3}, where, in the probabilistic view, $S$ represents the total choice probability, equal to 1, and noise tolerance $N$ denotes the probability of a single choice, equal to the reciprocal of the total number of admissible choices, including the \textit{null} choice. The surprise potential (see \autoref{Bayes}) represents an energy potential with higher energies (negative log probabilities) reflecting smaller probabilities. Competing neurons relax this potential over time, until one reaches the decision threshold $S_N$ \cite{Usher:2001}. Solving for time yields: 
\begin{equation}
\label{equation:Hicks2}
S(t)=S_0 e^{-t/b'}
\end{equation}
\begin{equation}
\label{equation:HicksError}
RT=b'\ln\!\left(\frac{S_0}{S_N}\right)
=
b'\ln 2
\log_2\!\left(\frac{S_0}{S_N}\right)
=
b
\log_2(n+1)
\end{equation}
\;

where $t$ denotes time, and, in the count ratio view, $S_0$ denotes the initial number of possible alternatives, including the case of no stimulus, $S_N=1$ the stopping condition corresponding to the single resolved choice, and $b$ the empirically determined time per decision bit. 

\subsection {Power Law of Practice}

 The third law of HCI, the Power Law of Practice \cite{Crossman:1959}, governs the learning curve that arises as users are learning a new task. Again, this is a learning task in which the prior is updated to the posterior. The theoretical basis for the Power Law of Practice was not explained: it is based on an empirical finding that was confirmed in many experiments \cite{Crossman:1959}. These suggest that the amount of time $T$ spent performing a trial of a task is a negative power function of the number of trials $x$ performed:
 \begin{equation}
\label{equation:PowerLaw}
T= a \cdot x^{-b} +  c
\end{equation}
\;
where $a$ and $b$ are empirically derived constants. Within the context of Interactive Inference, this Power Law is explained as a reduction in the KL divergence between prior and posterior that occurs when users are learning a task. This reduction is a function of the signal-to-noise ratio improving with every trial. In deep learning, the equivalent process is commonly formulated as minimizing negative log-likelihood \cite{bishop2023deep}:
\begin{equation}
\label{equation:NegativeLoglikihood}
I = a-b \cdot \log_2 \left( \frac{S}{N} \right)
\end{equation}
\;
where (a) and (b) are empirically derived constants. This equation describes the reduction in entropy, measured in bits, that accompanies an improvement in the task’s signal-to-noise ratio. Assuming a constant time per bit, it yields a logarithmic learning curve that approximates the Power Law over a finite range of trials.

\section{The Case of Unequal Variances}
There are some cases in which the simplification of the KL divergence to a square function of the SNR does not work. Consider the example of the distribution of socks that need to be placed in the drawer. We note that we could consider this task a compound task in which we sum the individual KL divergences of the movement of each individual sock. In this example, we are simply going to take the average of these, in an attempt to describe the overall relative entropy of the task with one KL metric.

We recognize that the surprise in this task is not just governed by the noise scale, size of the drawer or average distance of socks, but also by the number and distribution of socks in the room or house. This means that there may be a discrepancy between the size of the distribution of socks, our posterior, and the goal, our prior. This discrepancy also increases the KL divergence describing the surprise of the task. To calculate the KL divergence, we need to capture this difference. While the original integral does this for any distribution, including non-Gaussians, it is computationally expensive and not formulated in terms of SNR. There exists a more general function that is mathematically identical to \autoref{equation:KL} that describes the KL divergence in cases of Gaussian distributions with unequal variances \cite{Moulin:2014}: 
\begin{equation}
\label{equation:KL3_proof}
\begin{split}
\log_{2}\frac{P(s)}{G(s)}=\log_{2}\left(\frac{\sigma_{g}} {\sigma_{p}}\right)+\log_{2}\exp\left(-\frac{(s-\mu_{p})^{2}}{2\sigma_{p}^{2}}+\frac{(s-\mu_{g})^{2}}{2\sigma_{g}^{2}}\right) 
\\
KL \left(P(s)\:||\:G(s)\right) =\log_{2}\left(\frac{\sigma_{g}}{\sigma_{p}}\right)+\mathbb{E}_p \left[ \frac{(s-\mu_{g})^{2}}{2\ln\left(2\right)\sigma_{g}^{2}}\right] -\mathbb{E}_p \left[ \frac{(s-\mu_{p})^{2}}{2\ln\left(2\right)\sigma_{p}^{2}}\right]
\end{split}
\end{equation}
The expectation under $P(s)$ equals:
\begin{equation}
\begin{split}
\mathbb{E}_p\left[(s-a)^{2}\right]=(\mu_p-a)^{2}+\sigma_p^{2}
\\
\mathbb{E}_p \left[ \left(s-\mu_g\right)^{2} \right]=(\mu_p-\mu_g)^{2}+\sigma_p^{2}
\\
\mathbb{E}_p \left[ \left(s-\mu_p\right)^{2} \right]=(\mu_p-\mu_p)^{2}+\sigma_p^{2} = \sigma_p^2
\end{split}
\end{equation}
Substituting expectations:
\begin{equation}
\begin{split}
\label{equation:KL3}
KL \left(P(s)\:||\:G(s)\right)=\log_{2}\left(\frac{\sigma_{g}}{\sigma_{p}}\right)+\frac{\left(\mu_{p}-\mu_{g}\right)^{2}+\sigma_{p}^{2}}{2\ln\left(2\right)\sigma_{g}^{2}}-\frac{\sigma_{p}^{2}}{2\ln\left(2\right)\sigma_{p}^{2}}
\end{split}
\end{equation}
Here, $\mu_g$ and $\sigma_g$ signify the mean and standard deviation of the goal distribution. $\mu_p$ and $\sigma_p$ signify the mean and standard deviation of the progress distribution. The natural logarithms are constants, and are introduced by base conversion. Including these allows for an exact result in bits of relative entropy. The main difference between \autoref{equation:KL_SNR} and \autoref{equation:KL3} is the addition of the log variability ratio \cite{Senior2020} of goal and progress distributions. As such, \autoref{equation:KL3} models the \textit{mismatch} in variances as an increase in the KL divergence that causes the average entropy to increase, raising the entire surprise function. 
\subsection{Precision/Speed Trade-off}
Variance $\sigma_g^2$ represents the base noise scale of the scaled goal tolerance $zN$, while variance $\sigma_p^2$ describes the uncertainty of the progress distribution towards that goal. Together, these quantities capture a precision-speed trade-off. The relationship to Fitts’ Law is evident: when movement speed exceeds what the goal tolerances permit, the movement becomes suboptimal, either by increasing terminal error or by requiring corrective time. Stated more generally, when the user moves outside the tolerance limits for task outcome, the task will be executed sub-optimally. 

\section{Empirical Study}

We now turn to a case study, where we empirically examine whether our general model can  predict human performance in a particular situation. If it can, it provides evidence that the general model is worthy of further exploration.
In particular, we want to verify, via an empirical study, whether the above postulates could successfully model a task that does not yet have a known model, and function as a real-time measurement tool. We used our simplified KL divergence to measure the information-processing capacity of participants in a simulated target \textit{avoidance} task: maintaining distance while driving behind a car. Our attempt to develop a better model of a driver's capacity is more than an academic exercise, as such a model can help modern car interfaces better recognize and mitigate problems of driver distraction~\cite{Groeger_Understanding_2000}, by directly modeling cognitive load based on SNR.

\subsection{Study Literature Review}
In-depth analysis of crash data shows that nearly one third of all fatal and injury crashes involve driver inattention~\cite{wundersitz_driver_2019}.
Lee et al. identified a 30\% (310\textit{~ms}) increase in reaction time when a speech-based email system is used, with implications for safety~\cite{Lee_Speech_2001}.
Strayer et al. demonstrated that inattentional blindness can be caused by concurrent cell phone usage while driving, significantly increasing the risk of an accident~\cite{Strayer_Cognitive_2011}.
In a study involving adult drivers, Strayer and Drews observed an 18\% slower reaction time while conversing on a cell phone~\cite{Strayer_Profiles_2004}.
Systematic reviews of distracted driving, such as those by Simmons et al.~\cite{Simmons_Safety_2016, Simmons_2017} and Caird et al.~\cite{caird_2014} overwhelmingly demonstrate that tasks that require the drivers' visual attention, such as dialing,  texting, and in-car infotainment touch screen use generate more risk than tasks that do not require visual attention, as these direct the driver's attentive resources away from the primary task~\cite{niu_effects_2019}.
Researchers have shown particular interest in the car following task due to its prevalence in driving. Wolfe et al.'s \textit{information acquisition theory} modeled the information a driver acquires through peripheral vision and eye movements~\cite{wolfe_toward_2022}. 
Boer pioneered the use of a behavioral entropy measure as an index of workload, and suggested that information entropy (the average amount of information) is the basis for driving behaviour~\cite{boer_behavioral_2000}. 
Boer et al. used the sample entropy algorithm to gauge performance of drivers, as demonstrated through a driving simulator study~\cite{boer_steering_2005}. 
Senders did pioneering work on \textit{occlusion distance}: the distance a driver feels comfortable driving with their eyes shut~\cite{Senders_attentional_1967}. Kujala developed this metric further as a way to measure visual demand while driving~\cite{kujala_attentional_2016}.
Building on Wilde's \textit{risk homeostasis theory}~\cite{Wilde_theory_1982}, Lu et al. developed a \textit{desired safety margin} model as a quantified index of risk perception when car following~\cite{Lu_car_2013}. That model is based on the leading and following car speeds, and the relative distance between them.
Zhu et al. analyzed the driver's choice of headway (the time interval between two cars at a given speed) in various traffic conditions using a large database~\cite{zhu_car-following_2016}.
Wu et al. proposed a risk repulsion factor \textit{F} for car-following as determined by the time headway, relative speed, and safe spatial headway~\cite{wu_longitudinal_2021}. The driver's cognitive load has also been assessed and captured as a metric when car-following~\cite{summala_brake_2000}. This is important especially when secondary tasks consume some or all of that load. Subjective questionnaires are often carried out to collect the participants' assessment of task difficulty, risk, comfort and effort. For example, Lewis-Evans et al. identified a negative relationship between the time headway in a car following task and ratings of task difficulty in a simulator experiment~\cite{lewis-evans_thats_2010}. However, there are many issues with subjective assessments, including but not limited to bias caused by participants' preferences towards socially desirable responses~\cite{lajunen_can_2003}. Others have proposed various physiological measurements as indicators of the subject's mental effort, such as heart rate variability, electromyogram, and facial temperature~\cite{Mulder_Measurement_1992, deWaard_mental_1996, Kajiwara_Evaluation_2014}. Yet such portable physiological sensing solution are unreliable, while trajectory-based risk assessment methods are too complex~\cite{Petelczyc_Impact_2020}. 

\subsection{KL Divergence of Overlap}
\label{sec:model.prob}

\begin{figure*}[t]
\includegraphics[width=\linewidth]{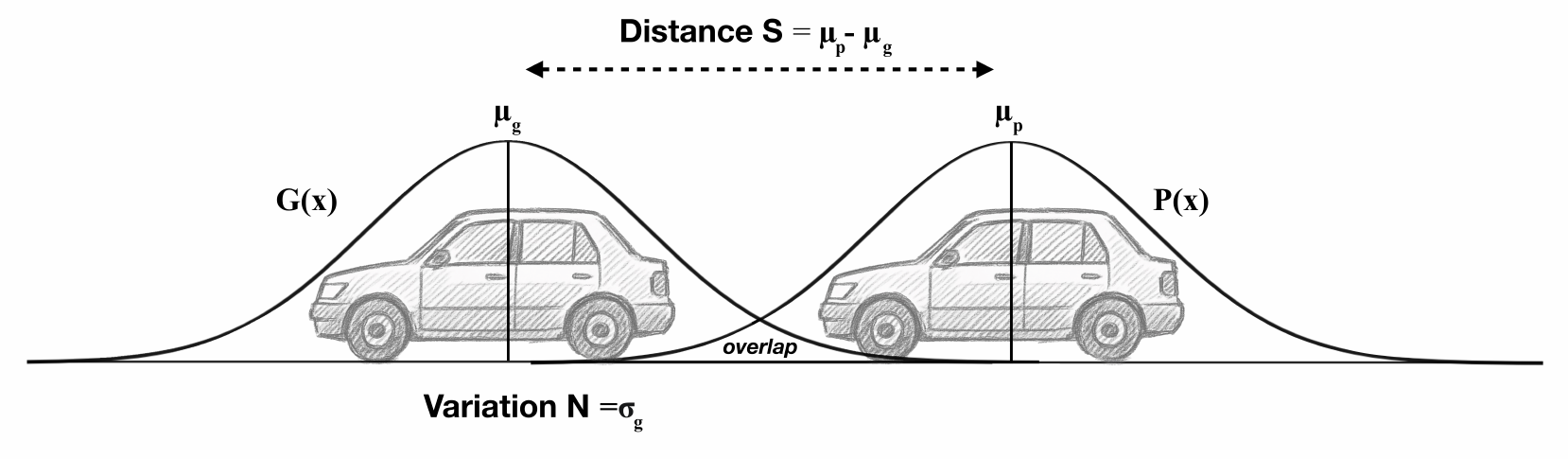}
\caption{Two cars have normal distributions \(G(x)\) and \(P(x)\) representing uncertainty in their positions. Their overlap determines the modeled probability of a collision. The negative logarithm of this probability tracks the KL divergence: the number of bits distinguishing the two distributions.}
\label{fig:car.following2}
\Description[One car follows the other along a horizontal axis.]{One car follows the other along a horizontal axis, labeled as P(x) and Q(x), with a distance of S. The front car's location has a variation of N along the horizontal axis.}
\end {figure*}

We used our framework to model the car following task probabilistically, as illustrated in \autoref{fig:car.following2}. The task is represented using two normal distributions that describe the uncertainty in the lead and following car’s position. In our model, these are the goal distribution (of the front car) and a progress distribution (of the following car). The task for the driver of the following car is to ensure the progress distribution avoids overlap with the goal distribution. In a goal-avoidance task, surprise increases as the progress distribution approaches the forbidden goal distribution. We therefore invert the goal-directed normalized error S/N towards goal-avoidance, yielding N/S. The overlap between the two distributions follows a KL divergence that indicates the amount of bits of processing by the following driver to perform this task. Our simplified KL divergence models this as follows:
\begin{equation}
\label{equation:Car_SNR}
U= \frac{1}{2\ln2} \cdot \left( \frac{\sigma_g}{\mu_p-\mu_g} \right)^2 = \frac{1}{2\ln2} \cdot \left( \frac{N}{S} \right)^2
\end{equation}
\;

where $\mu_g$ is the mean of the goal distribution and $\mu_p$ is the progress distribution. Recognize that what matters here is  the mean relative distance between cars $(\mu_p-\mu_g)$ as it relates to variability in the goal, the deceleration behaviour of the front car $\sigma_g$ that needs matching. Note also that since we are avoiding rather than meeting the goal, we need to take the reciprocal of the signal-to-noise ratio, i.e., the noise becomes the signal. 

$U$ is the amount of Bayesian surprise, in bits, that the driver would need to process in a car following task. However, recall that the capacity $C$ of the driver's ability to process this surprise is only logarithmic. We can establish parameters $a$ and $b$ of this logarithm for target \textit{avoidance} by measuring the amount of information gained between the start and end of each trial using \autoref{equation:Car_SNR} and by performing a regression on the result (we discuss how we operationalized this in the \textit{Results} section, see \autoref{equation:InformationGain}):

\begin{equation}
\label{equation:Car_Capacity}
C= a + b \cdot \log_2\left(\frac{z\sigma_g}{\mu_p-\mu_g} \right) = a+b \cdot \log_2 \left( \frac{zN}{S} \right)
\end{equation}
\;

\subsection{Task}
\label{sec:study}
We designed a controlled car following task in a driving simulator to collect driving data in target avoidance conditions with varying z-scaled noise-to-signal ratios. Our goal was to model the driver's information load using our equations by quantifying the information required to maintain a following distance \(S\) from a lead vehicle with a variation of \(zN\). We focused our analysis on deceleration events that could lead to a collision.

$S$ and $zN$ were varied between conditions. 
As in real driving, $zN$ was not static within a condition: the lead car accelerated and decelerated to different speeds, which added a normally distributed noise $zN$ to each distance condition, at different levels of $z$. 
This allowed us to capture participants' handling of the car in a range of driving difficulty levels defined by the constituting $zN/S$ ratios. 
We hypothesized that driver's information gain during decelerations would be logarithmic with the z-scaled noise-to-signal ratio, with a mean capacity $b$ in bits of surprise per bit of difficulty. Because information gain increases logarithmically while generated surprise increases quadratically, the generated information will exceed the driver’s processing capacity beyond the mean-capacity point \(b\). We hypothesized that the resulting residual surprise would follow the KL-divergence form in \autoref{equation:Car_SNR}.

\subsection{Participants}
We recruited fifteen participants (5 female, 10 male) from our organization, with a mean participant age of 27. Four of the participants had low driving experience (one year or less).

\subsection{Experimental Setup}
\label{sec:study.setup}
\begin{figure}[t]
\includegraphics[width=\linewidth]{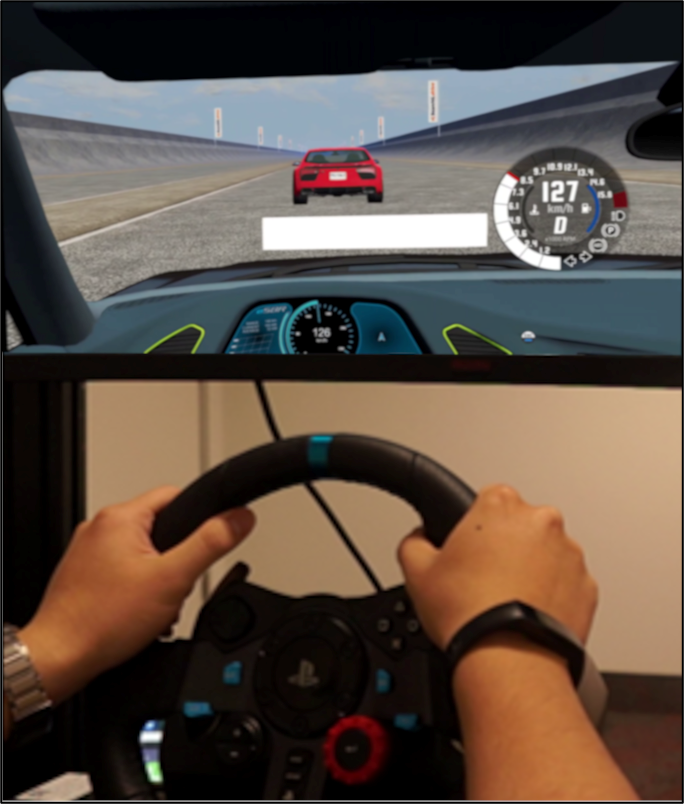} 
\caption{Participant's experimental setup. The white bar in the center of the screen helped participants perceive the correct distance $S$. They needed to match the size of this bar with the lead car's bumper while driving. Here, it is wider than the lead car's bumper indicating that the participant needs to accelerate.}
\label{fig:car.carfollowingui}
\Description[Two hands hold a simulator steering wheel in front of a monitor.]{Two hands hold a simulator steering wheel in front of a monitor, which displays an in-car field-of-view, a white bar in the middle, a speedometer on the right, and a leading car's rear bumper. The white bar is between the leading car and the user's car. }
\end {figure}

We presented a simulated driving task with simulated road conditions using BeamNG~\cite{beamng_2013}, a car simulator video game chosen for its realism, physics engine, low cost, and programmability. Each participant driver was seated in a simulator in front of a curved 49-inch monitor with a 32:9 screen ratio. 
The monitor was positioned 0.6~m from the participant at eye level to provide an immersive simulator view of a car dashboard and windscreen (see \autoref{fig:car.carfollowingui}). 
A Logitech G29 steering wheel and pedal unit with three pedals were mounted under the simulator rig. 
Participants were asked to adjust their seat position so that the steering wheel and accelerator/brake pedals were at a comfortable distance. The pedal clutch was not used. Every participant used the same custom car model with the same physics characteristics in BeamNG.

We constituted our conditions through custom scripts in BeamNG. 
We also wrote a Python program to interface with BeamNG to administer the study and to capture driving data at a 450~Hz sampling rate: vehicle positions and speeds of the driver's car and lead car, pedal data, and collision data. 
Our custom Python scripts filtered and analyzed the data.

During the experiment, participants had a clear view of the lead car and road through their simulated windscreen. Because a 2D simulator does not provide 3D depth cues, participants were shown a floating fixed-width rectangular bar in the center of their windscreen, positioned below the bumper of the lead car. That bar's width reflected what the perceived width of the lead car's bumper should be at the optimal following distance. The driver's task was to maintain an optimal following distance, where the driver would try to match the bar's size to the perceived size of the bumper of the lead car. For example, in \autoref{fig:car.carfollowingui}, the bar is wider than the bumper, meaning the driver should move closer to the lead car. 

\subsection{Experiment Design}
We used a within-subject experiment design where the task was to follow the lead car at a specified distance $S$. We then varied the speed of the lead car with variation $zN$. Each participant experienced 12 different task conditions with two factors $S$ and $zN$.
\begin{itemize}[align=left, leftmargin=*] 
\item Factor $S$ comprised four levels of fixed distances to the lead car (2.84~m, 4.84~m, 6.84~m, and 8.84~m). 
\item Factor $zN$ comprised three levels centered around a mean lead car speed of 100~km/h. Each level represented  a 4.1$\sigma$ of a normal distribution of varying distances to the lead car, as generated by increasing the range of the lead car's allowable speeds (1.93~m: 100±12~km/h; 4.19~m: 100±18~km/h; and 7.21~m: 100±24~km/h). That is, the levels represented by factor $z$ created increasing variability and thus noise in how the car in front accelerated and decelerated. To clarify these numbers, at level 1 the speed was allowed to vary between 88 and 112~km/h (i.e., 100±12~km/h), where 96\% of the variation in distance was captured by 1.93~m. 
\end{itemize}
To constitute each distribution, for each level of $zN$, the lead car randomly decelerated 20 times and randomly accelerated 20 times, although always keeping within the level's range of speed values. Each trial within a level of $zN$ comprised a new speed adjustment, which occurred approximately every two seconds.
We randomized the presentation order of conditions. We also randomized the presentation order of acceleration or deceleration trials, with accelerations never occurring within a deceleration trial or vice versa. 

We measured the instantaneous noise-to-signal ratio (change in distance by lead car $\Delta S$, i.e., some sample of $N$; mean distance to lead car $\overline{S}$, per trial), and pedal activity by the driver.
As will be detailed further, our dependent variables were the amount of information gained during a trial ($\Delta$ KL divergence), and error (collisions). Together, the 40 trials created a condition simulating a continuous driving task of 2 minutes. 

\subsection{Procedure}
Participants were instructed to hold but not use the steering wheel. 
They were instructed to only use the brake and accelerator pedals to maintain the distance $S$, as the lead car varied its speed, by matching the bar with the bumper of the lead car. 
The road simulation was straight and without obstacles, allowing cars to travel in a straight line. 
Participants were given a 5 minute training task to familiarize themselves with the controls and task.
Each condition began with both cars automatically accelerating to 100~km/h, with the specified distance $S$. 
Participants were then given a visual alert to take control of the car. 
After each condition ended, we asked the participants to rate the driving difficulty of that condition using a 7-point Likert scale~\cite{lewis-evans_thats_2010}. 
Trials that resulted in collision were repeated, with any collisions only counting once per trial, per participant.

\subsection{Ecological Validity}
Our task was designed such that participants would be in a position to experience error. A choice for longer distances and lower speeds would mean the chances of a collision would be too low to be measurable and compared with the driver's remaining capacity. This meant having relatively low headways at high absolute speeds that could be criticized for their limited ecological validity. We note however that the important parameter is not the absolute speed at which the cars are driving, but the variation in their average relative distance over time. This applies equally to parking a car with respect to a solid barrier at low speeds, as it does to avoiding a lead car at 100 km/h. We also chose for the simulator rendering to have most environmental cues removed, making it difficult for participants to estimate an absolute speed. This choice made the experiment more generalizable by avoiding potential confounding factors generated by environmental renderings. The choice not to have participants steer was inspired by the potential for introducing confounding variables as well. 
\section{Results}
\label{sec:results}
We first describe our data analysis, and then present the results of our empirical evaluation of the models.

\subsection{Measurement and Data Analysis}
\label{sec:measurement}
In our experiment, lead car accelerations never resulted in a collision. Given our interest in the relationship between target avoidance and the risk of collision (error), we constrained our analysis to trials where participants responded to a deceleration of the lead car by braking or by releasing the accelerator pedal. We used 3 different levels of tolerance for noise-to-signal measurements in our analysis. 
Each used a different confidence interval $z$ for defining $zN$.
\begin{enumerate}[label=(\arabic*), align=left, leftmargin=*] 
    \item At the condition level, $zN$ was defined as 4.1$\sigma$ of the distribution of relative backward distances traveled by the lead car during trials in that condition; 
    \item At the trial level,  $zN$ was defined by the actual relative backward distance traveled by the lead car during each trial; and 
    \item Within each trial, $z=1$, and $N$ was defined by the standard deviation $(\sigma)$ of a one-second time window of instantaneous distance measurements, as detailed below.
\end{enumerate}

\subsubsection{Empirical Factors Noise (zN) and Signal (S)}
\label{measuringnoisesignal}
Although we used predetermined levels of factor $S$, we administered factor $zN$ by altering the speed of the lead vehicle, which was affected by the physics engine. 
As such, we measured the actual negative relative distance traveled by the lead car ($\Delta S$) per trial, from the moment of deceleration of the lead car to the lead car reaching the desired speed. 
In line with Welford's statistical analysis of Fitts' law target width $W$~\cite{Welford_fundamentals_1968}, we then calculated $zN$ as 4.1$\sigma$ of the distribution of the $\Delta S$s actually administered per level of condition $zN$. 

\subsubsection{Empirical Noise-to-Signal Ratio}
\label{noisesignal}
As each trial involved a braking event that resulted in processing of information given an individually administered $\Delta S$, many similar $zN/S$ ratios were created across conditions. 
We used these more detailed empirical observations of noise-to-signal ratios for our log-linear regression analysis, pooled across all trials. 
Here, our definition of noise-to-signal ratio was $N=\Delta S$ per trial divided by the measured mean $S$ ($\overline{S}$) per trial, yielding the base metric $\Delta S/\overline{S}$. 
We also used this definition for our regression analysis of normalized error frequency, as this yielded directly comparable results of terminal error vs. information gain, in bits. This definition was also used for our Index of Difficulty (see \autoref{equation:W_A_ID}).
Note that the only effect of the use of sampled $\Delta S$ values instead of some confidence interval $zN$ is a scaling effect $z$ on $N/S$ ratios.

\subsubsection{Measuring Information Processed}
Our first dependent variable was the amount of information processed by the driver during decelerations. To measure the instantaneous amount of information processed within a trial, we used a more detailed 1-second rolling time window to sample the distribution of distances $S$ to the lead car at 450~Hz. 
Our base definition of noise $N$ here was the standard deviation ($\sigma$) of the distribution of $S$ obtained in this time window, while our definition of signal $S$ was the mean $S$ ($\overline{S}$) of this time window. 
First, the noise-to-signal ratio was sampled using this type of one second window at the beginning of each trial, ending at the moment the driver responded to the stimulus. We then calculated the amount of surprise in the stimulus by entering this first noise-to-signal ratio into \autoref{equation:Car_SNR}. 
A one-second window again sampled the noise-to-signal ratio at the end of each trial, prior to (i.e., ending at) a new deceleration or acceleration event. We calculated the remaining amount of surprise by entering this second noise-to-signal ratio into \autoref{equation:Car_SNR}.

Subtracting the remaining amount of surprise from the amount of surprise in the stimulus gave us the amount of information gained $I_b$ by the driver during each deceleration trial:
\begin{equation}
\label{equation:InformationGain}
I_b=U_{\mathrm{initial}} -U_{\mathrm{final}}=
\frac{1}{2\ln2}\left[\left(\frac{\Delta S_{\mathrm{initial}}}{\overline{S}_{\mathrm{initial}}}\right)^2-\left(\frac{\Delta S_{\mathrm{final}}}{\overline{S}_{\mathrm{final}}}\right)^2\right]
\end{equation}
We subsequently aggregated measurements across all conditions, sorting results using $\Delta S/\overline{S}$, and binning measurements that were within 0.05 units of $\Delta S/\overline{S}$. This $I_b$ was then fitted using the logarithmic-capacity model from \autoref{equation:Car_Capacity}:
\begin{equation}
\widehat{I}_b=\hat{a}+\hat{b} \cdot \log_2 \left( \frac{\Delta S}{\overline{S}} \right)
\end{equation}

\subsubsection{Measuring Error}
Our second dependent variable was terminal error. Terminal error was defined, across conditions, as the occurrence of a collision with the lead car. 
All trials that ended in error were repeated. 
Since any instantaneous measure of $N/S$ during a collision would yield infinity during the 1 second period leading up to the collision, we used the $\Delta S/\overline{S}$ of the repeat of the trial as our signal-to-noise metric. 
This allowed comparison of error and capacity metrics using the same scale. We could not establish any measure for residual Bayesian surprise for trials that ended in collision. We could, however, convert the normalized frequency of error into bits of information in a manner that provided insights into the relationship between driver capacity and error rates. 
After binning errors within 0.5 $\Delta S/\overline{S}$, we calculated the normalized frequency of error $p_{\mathrm{err}}$ by dividing the number of errors per bin by the total number of errors across all conditions. 
We then converted the complement of this probability to bits $E_{b}$ (the surprise of not having a collision) using a negative log probability transform~\cite{shannon_mathematical_1948} and regressed this against the KL in \autoref{equation:Car_SNR}:
 \begin{equation}
\label{equation:Bits}
E_b=-\log_2\left(1-p_{\mathrm{err}}\left(\frac{\Delta S}{\overline S}\right)\right), \; \widehat{E}_b = \hat{\alpha}+\frac{\hat{\beta}}{2\ln 2}\left(\frac{\Delta S}{\overline S}\right)^2 
\end{equation}
\begin{table}
  \caption{Mean reduction in KL divergence (s.e.) representing information gained in bits, for levels of factors $zN$ and $S$.}
  \label{table:first}
  \Description[A table of processed information in bits of 4 levels S against 3 levels N.]{A table of information gained in bits of 4 levels S per column (2.84/4.84/6.84/8.84) against 3 levels N per row (1.93/4.19/7.20). Processed bits range from 0.007, of N 8.84 m x S 1.93 m, to 0.123, of W 2.84 m x A 7.20 m.}
  \resizebox{\columnwidth}{!}{
  \begin{tabular}{ccccc}
    \multicolumn{5}{c}{\textbf{S}} \\
    \toprule
    \textbf{zN} & \textbf{2.84 m} & \textbf{4.84 m} & \textbf{6.84 m} & \textbf{8.84 m}\\
    \midrule
    \textbf{1.93 m} & 0.079 (0.050) & 0.043 (0.029) & 0.022 (0.014) & 0.007 (0.014) \\
    \textbf{4.19 m} & 0.115 (0.065) & 0.065 (0.029) & 0.036 (0.022) & 0.029 (0.029) \\
    \textbf{7.20 m} & 0.123 (0.065) & 0.072 (0.022) & 0.058 (0.029) & 0.036 (0.014) \\
    \bottomrule
  \end{tabular}}
\end{table}
\subsection{Effects of Factors \textit{zN} and \textit{S}}
\autoref{table:first} shows the mean reductions in KL divergence representing relative information gained, against levels of factors $zN$ and $S$. Since the residuals in our ANOVA model of the dependent variable differed significantly from a normal distribution (KS test, p<0.05), we used a non-parametric Friedman test. The Friedman test showed highly significant effects of both factors $zN$ ($\chi^2(2)=17.73, p<0.001$) and $S$ ($\chi^2(3)=40.84, p<0.001$) on the number of bits processed by the driver (note that a two-way ANOVA produced the same results).
\begin{figure*}[t]
\includegraphics[trim=.2cm .1cm .8cm 0,clip,width=\linewidth]
{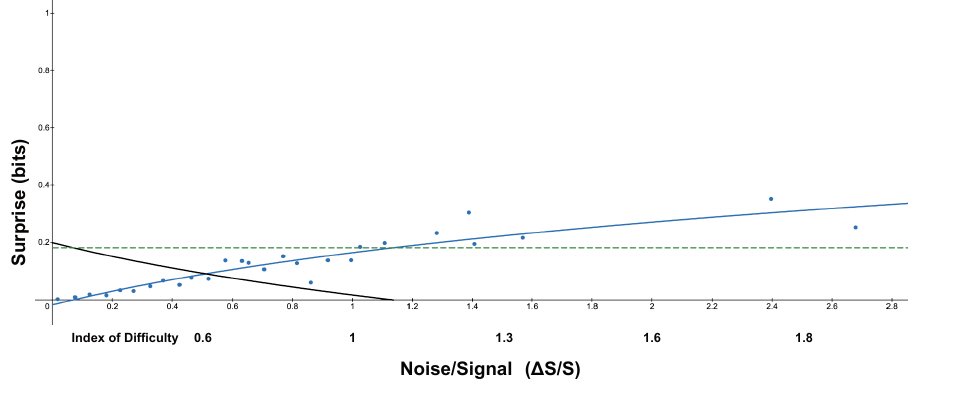}
\caption{ Driver capacity (relative information gain) plotted against noise-to-signal ratio ($\Delta S/\overline{S}$, with index of difficulty (ID) on the logarithmic scale below the x-axis). Capacity followed a logarithmic function (blue line), with a slope of 0.18 bits (green dash). Black solid line shows remaining capacity, and suggests participants ran out of capacity near an ID of 1.1 bit.}
\label{fig:car.infoload}
\Description[A curve plot of three lines, and a series of data dots along the rising logarithmic line.]{A curve plot of three lines: 1) Line 1 shows driver information capacity following a logarithmic function, 2) Line 2 plots mean capacity of 0.18 bits in parallel to x axis, and 3) Line 3 shows a dropping line indicating remaining capacity. Lines intersect around 1.1 bit. The x-axis is noise-to-signal ratio on a logarithmic scale. The data dots spread along Line 1, the driver information capacity.}
\end {figure*}
\subsection{Car Following Model}
\label{carmodel}
Log-linear regression of binned KL divergence reductions against noise-to-signal measurements $\Delta S/\overline{S}$ yielded a logarithmic relationship $(r^2=0.86, F(1,26)=164.2, p<0.001)$:
\begin{equation}
\label{equation:W_A_Result}
I_b = -0.01+0.18 \cdot \log_2 \left( \frac{\Delta S}{\overline{S}} + 1 \right)
\end{equation}
\autoref{fig:car.infoload} shows this relationship between driver capacity (the information gain or processed Bayesian surprise) and aggregated signal-to-noise measurements on a $\Delta S/\overline{S}$ scale across conditions, as well as on a logarithmic scale (below the x-axis). 
Mean capacity, the empirically observed capacity coefficient $\hat{b}$, is visualized as a green dashed line. Subtracting the information processed by the driver from mean capacity yields spare driver capacity (black solid line). This can be interpreted as a metric for the cognitive/sensorimotor spare capacity of the driver. The logarithmic scale at the bottom of \autoref{fig:car.infoload} shows the index of difficulty (ID) of this task, in bits:
\begin{equation}
\label{equation:W_A_ID}
ID = \log_2 \left(\frac{\Delta S}{\overline S} + 1 \right)
\end{equation}

The mean capacity coefficient of our car following task was measured at $b=0.18$ bits of surprise per bit of task difficulty. 
The y axis indicates bits of processed surprise, or information gained, while the x axis indicates noise-to-signal ratio $\Delta$S/$\overline{S}$, or bits of $ID$ if the bottom log scale is used.
Note that the only difference in plotting \autoref{fig:car.infoload} on a log-linear scale against $ID$ is that lines would be straight, with the bottom scale as the primary x-axis. 

\subsection{Error (Collision) Model}
\label{sec:error}
Regression of error surprise metric $E_b$ in \autoref{equation:Bits} (surprise of not having a collision, in bits) yielded a significant relationship with the KL divergence metric ($r^2=0.94, F(1,5)=80.5, p<0.001)$:
\begin{equation}
\label{equation:Error}
E_b=
0.03+0.16\left[
\frac{1}{2\ln 2}
\left(\frac{\Delta S}{\bar S}\right)^2
\right]
\end{equation}

\begin{figure*}[t]
\includegraphics[trim=.2cm .1cm .8cm 0,clip,width=\linewidth]{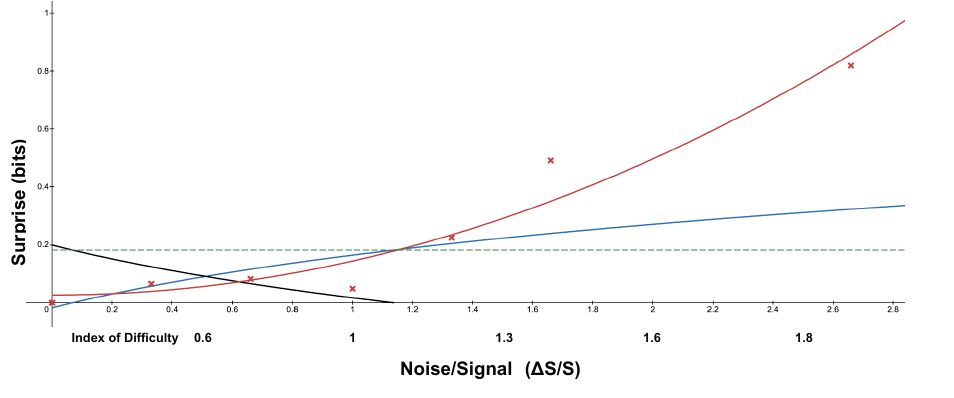}
\caption{Error surprise (surprise of not having a collision, red crosses, in bits) plotted against noise-to-signal ratio ($\Delta S/\overline{S}$, with index of difficulty (ID) shown as a logarithmic scale below the x-axis). Error surprise followed a KL divergence form (red line). Blue line shows driver capacity and black line spare capacity. Lines cross near an ID of 1.1 bit, marking overloading of participants.}
\label{fig:car.error}
\Description[A curve plot of three lines, and a series of data dots along the rising power line.]{A curve plot of three lines: 1) Line 1 shows an error curve following a KL divergence with data dots along it, 2) Line 2 shows a driver capacity line following a logarithmic curve, and 3) Line 3 shows a dropping spare capacity curve. Lines intersect around 1.1 bit. The x-axis is noise-to-signal ratio, with logarithmic scale.}
\end{figure*}

\autoref{fig:car.error} shows the relationship between terminal error (red solid) and aggregated signal-to-noise measurements on a $\Delta S/\overline{S}$ scale across conditions, as well as on a logarithmic scale ($ID$, bottom), compared with driver capacity (blue solid) and remaining capacity (black solid). 

The total number of collisions across all trials was 90. Error rates below or equal to a noise-to-signal ratio or $ID$ of 1 trended, on average, slightly below the expected 4\% (at 0.05 bits or 3.4\%, between 2 and 4 collisions across trials), while those above 1 averaged 0.38 bits (23\%), with a maximum of 0.82 bits (43\% or 39 collisions across trials) at $ID=1.85$.

\subsection{Qualitative Results}
\begin{figure*}[t]
\includegraphics[width=\linewidth]{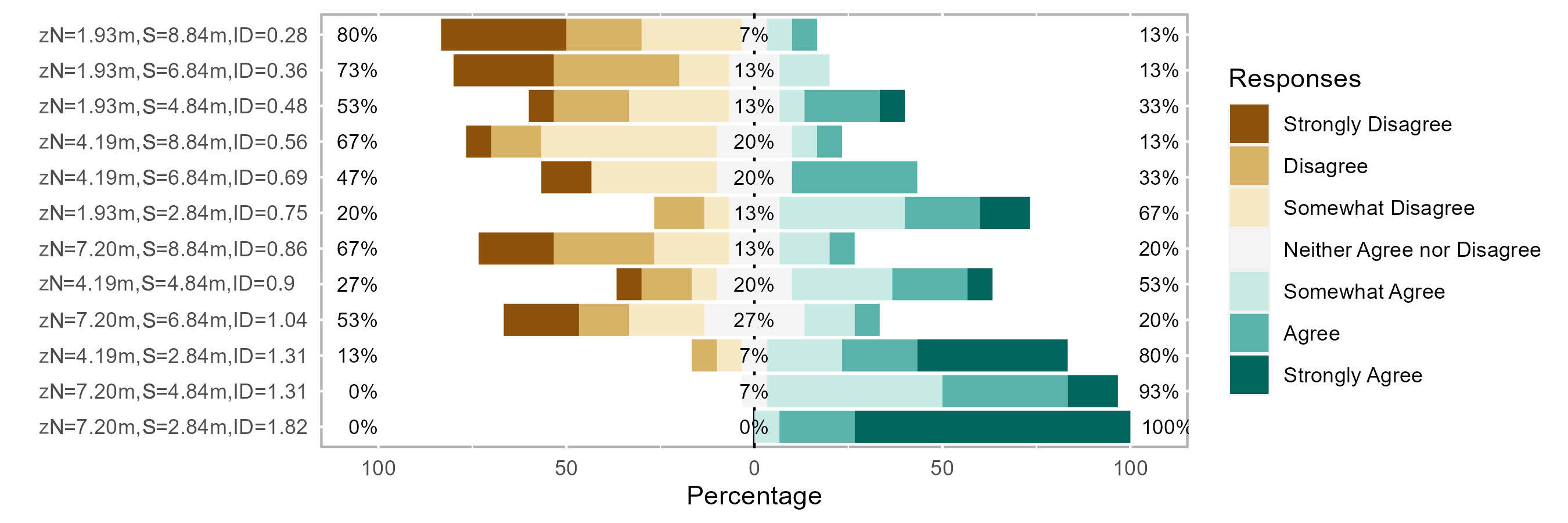}
\caption{Perceived task difficulty for each condition: participants' responses to the question ``This driving task was difficult'' on a scale of 1 (Strongly Disagree) to 7 (Strongly Agree).}
\label{fig:car.perceiveddifficulty}
\Description[A bar chart of all conditions' subjective ratings aligned vertically by the middle-point choice of “Neither Agree nor Disagree”.]{A bar chart of all conditions' subjective ratings (7-point scale) aligned vertically by the middle-point (level-4) choice of “Neither Agree nor Disagree”. In each row, the choices are sorted by the rating from 1 (“Strongly Disagree”) to 7 (“Strongly Agree”), left to right. For the easiest condition (row 1), the range of the user's rating is -0.80--0.13. For the hardest condition (row 12), the range of the user's rating is 0.00--1.00. }
\end {figure*}

\autoref{fig:car.perceiveddifficulty} shows participants' questionnaire responses ranking difficulty per condition $zN/S$, sorted by $ID$. 
These represent mean responses to the question ``This driving task was difficult'', on a scale from 1 (Strongly Disagree) to 7 (Strongly Agree)~\cite{lewis-evans_thats_2010}. 
A Friedman non-parametric analysis of variance showed participants rated the car following task as significantly more difficult with increasing $ID$ ($\chi^2(11)= 116.4, p<0.0001$). 
Note that all participants found the task to be easy to interpret and execute, while none reported distraction by the floating bar graphic during the task. 

\section{Discussion}
We will now discuss the results from our empirical study, relate them to our Interactive Inference model, and provide some suggestions for future directions.
\subsection{Empirical Study}
The above results are a first step towards empirical validation of both our capacity and terminal error models. Results show a good-to-excellent fit between predicted and observed models of the task. Driver information processing capacity during deceleration events in a simulated car following task can be measured in real time using the simplified KL divergence of the noise-to-signal ratio of the distance to the car followed, over a time window. Our hypotheses that both distance $S$ and its variation $zN$ to the lead car significantly affect the amount of information gain by the driver during such events were confirmed. 

Error frequencies also appear to independently confirm the effect of $zN/S$ ratios on driver information gain when car-following. As expected, the (negative log of the complement of) normalized error frequency followed a KL divergence. When driver capacity exceeded mean capacity, surprise of not having a collision appeared to increase with the square of the noise-to-signal ratio. We attribute these trends to residual Bayesian surprise, generated by high noise-to-signal ratios, that is not processed by the driver. Note that we used mean capacity as our capacity limit because the theoretical maximum channel capacity is, in practice, never reached without significant terminal error \cite{shannon_mathematical_1948}. Interestingly, the independently measured error surprise curve intersects both mean capacity and the logarithmic capacity curve at approximately \(1.1\) bits of task difficulty. This suggests that capacity is exhausted when the noise becomes comparable to the error signal, making reliable discrimination difficult. Our results suggest that the mean relative information gain was low, with an estimated mean capacity of \(0.18\) bits of surprise processed per bit of task difficulty. One possible explanation is that information processing was constrained by delays in the visuomotor control loop linking visual perception to foot pedal actuation.

Qualitative ratings by participants were in line with the above. They rated the task as more difficult with higher noise-to-signal ratios (and thus $ID$s), albeit with bipolar ratings for middle $ID$s. We attribute this to participants perceiving distant targets $(S>4.8~m)$ as easier to process than close targets with similar signal-to-noise ratios, perhaps due to the 2D nature of our simulator setup. We again note that our results are limited to modeling driver responses to lead car decelerations in a simulated car following task using specific road conditions, car physics, and a sample set of participants.
\subsection{Interactive Inference Model and Future Directions}
Our results support the Interactive Inference prediction that information processing capacity and terminal error can be expressed within a common information-theoretic framework. The fitted capacity coefficient is analogous to bandwidth in the Shannon--Hartley theorem and represents the maximum average information gain attainable without terminal error. Subtracting observed information gain from this capacity yields a cognitive/sensorimotor spare-capacity metric. As this metric approaches zero, error surprise begins to increase according to the quadratic dependence predicted by the equal-variance Gaussian KL-divergence model. This suggests that terminal error may be understood as residual surprise arising when task demands exceed available capacity. 

A direct empirical test of this spare-capacity interpretation remains a direction for future work. Under this interpretation, the proportion of information gain relative to the mean capacity corresponds to the cognitive/sensorimotor load imposed by the task, whereas the remaining proportion corresponds to the available cognitive/sensorimotor spare capacity. By correlating empirical results with Heart Rate Variability (HRV) data and other measures of cognitive load evidence for this relationship could be examined. One important topic of investigation is to determine if cognitive load could be measured during tasks, in real time. E.g., users could complete a task modeled via our method while wearing a heart rate monitor. HRV should be lower during moments in the task where there is substantial residual surprise. A correlation between HRV and cognitive load may suggest that residual Bayesian surprise also tracks stress levels in participants. The relationship between our models and experienced stress levels of participants during tasks is therefore an interesting area of future study. E.g., users could complete a task modeled via our method and then fill out a questionnaire on stress levels. Residual surprise metrics should correlate with questionnaire results if our hypothesis is correct. Because our metrics are in bits, rather than time, it should be possible to compare both capacity and error between different tasks. Analyzing different tasks for a signal-to-noise ratio parameter that can be measured, and comparing results from empirical investigations of the relationship between SNR and both capacity and error, would further evidence this assertion. E.g., users could complete different tasks while wearing HRV monitors. Spare capacity metrics should correlate not just with HRV data, but also between tasks, if our hypothesis is correct. We consider such empirical work a future direction.

\section{Limitations}
This work presents an exploratory framework inspired by Active Inference principles that does not require implementation of its full mathematical machinery (such as model evidence and expected free energy). Our use of the term Bayesian surprise refers specifically to the KL divergence between our operationally-defined goal and progress distributions. In addition, the connections we draw to Fitts' Law, Hick's Law, and the Power Law are observational. Finally, our empirical validation is limited to a single simulated driving task. Further research is needed to test the generalization of this approach to other interaction contexts.

\section{Conclusions}
Neuromorphic HCI aims to improve user interfaces by exploring how concepts from neuroscience might inform design. Interactive Inference offers a simplified interpretation of the neuroscience theory of Active Inference for user interface design. It postulates that users perform Bayesian inference on current progress and goal distributions to learn tasks and predict their next action. The Bayesian surprise (relative entropy) between goal and progress distributions can be modeled as a square function. However, human capacity to process Bayesian surprise follows a logarithmic function. This explains why terminal errors occur when task demands exceed the user's available processing capacity. We believe our model may allow the quantitative analysis of performance and terminal error in a single framework that allows estimation of the cognitive load of a task. We showed how three basic laws of HCI, Hick's Law, Fitts' Law and the Power Law of Practice can be derived from Interactive Inference. We evaluated the model in an empirical study. In our simulated car-following task, driver capacity (information gain) during braking events was linearly related, with a good fit, to the logarithm of the ratio between following distance and \(z\)-scaled noise, plus one, expressed in bits. We found a significant effect of both distance $S$ and variation $zN$ on capacity, with a mean driver capacity of 0.18 bits of surprise per bit of task difficulty (as defined by $log_2(zN/S + 1))$. Participants' qualitative ratings followed the same trend, with significantly higher ratings of difficulty with higher $zN/S$ ratios. We used our model to derive a spare processing capacity for the driver, by subtracting the log-linear curve from mean capacity, in bits. We also fit terminal error (collision) data, with excellent fit, to the square of the signal-to-noise ratio of the distance to the lead car, in bits. 
We hypothesize that collisions are due to residual surprise when driver capacity is exceeded. Overall, the empirical results provide support for our Interactive Inference model of HCI. While car-following is just one case study, it provides a framework for conducting further empirical research in Interactive Inference, and evidence that such research is well worth doing.  We invite other researchers to test and apply this model to other tasks, to see how well these models predict human performance, and to confirm or refute the general applicability of Neuromorphic HCI in general, and Interactive Inference in particular.
\section{Acknowledgements}
We would like to thank Jin Tang, Sean Braley, Calvin Rubens, Lili Shen and Rodney Etienne for their support.
\bibliographystyle{ACM-Reference-Format}
\balance
\bibliography{references}
\newpage

\clearpage
\onecolumn

\section*{Errata and Clarifications}
\textbf{This version contains the following corrections and clarifications to Chapter 57, `Interactive Inference: A Neuromorphic Theory of
Human--Computer Interaction,' in J. A. Jacko (Ed.),
\textit{The Human--Computer Interaction Handbook}
(4th ed.), CRC Press, 2026.}

\;
\;

\begin{enumerate}[leftmargin=14pt,itemsep=1em]
\item This version includes an abstract, minor spelling and grammatical improvements, with Canadian-English spelling.

\item Corrected definition of KL divergence in Equation 57.7. 

\item Clarified the definitions of Bayesian surprise, task-outcome scale,
    normalized error, task difficulty, noise scale \(N\), tolerance factor
    \(z\), and terminal execution error, and applied these definitions
    consistently throughout the chapter. 
\item Introduced explicit equations for information gain as the reduction in surprise, the corresponding logarithmic-capacity regression, and the empirical terminal-error surprise model, where these relationships were previously described only in prose or with scaling constants absorbed into fitted coefficients.
\item Corrected the general definition of KL divergence by removing the
    empirical coefficient \(\beta\). The coefficient \(\beta\) is now used only as a scaling coefficient in the empirical terminal-error model. For equal-variance Gaussian distributions:
\begin{equation*}
        D_{\mathrm{KL}}
        = \frac{1}{2\ln 2}\left(\frac{S}{N}\right)^2
\end{equation*}
    
    \item In the Results section, scaled the measured KL divergence by
    \(\frac{1}{2\ln 2}\). The original measurements were based on
    \(\left(\frac{\Delta S}{\overline{S}}\right)^2\). This rescaling does not
    change the statistical tests, goodness-of-fit results, or conclusions.

    \item This rescaling does change the value of the mean-capacity coefficient (to \(b=0.18\) bits of surprise per bit of task difficulty),
    the values in Table~57.1, and blue and green lines in Figures 57.5 and 57.6. The blue and green lines now correctly intersect error surprise (red line) at 1.1 bits.

    \item Corrected the fitted terminal-error equation so that the empirical
    coefficient \(\beta\) no longer absorbs the factor \(\frac{1}{2\ln 2}\). Note that this does not actually change its value:
    \begin{equation*}
        E_b
        = 0.03
        + 0.16\left[
            \frac{1}{2\ln 2}
            \left(\frac{\Delta S}{\overline{S}}\right)^2
        \right]
    \end{equation*}

    \item Corrected label of first row in Table 57.1 from N to zN.

    \item Corrected the Kolmogorov--Smirnov test statement from \(p>0.05\) to
    \(p<0.05\).

    \item Added Acknowledgements.
\end{enumerate}


\end{document}